\documentclass[aps,prd,twocolumn, showpacs, nofootinbib]{revtex4-1}
\usepackage{bm}
\usepackage{amsmath}
\usepackage{amsfonts}          
\usepackage{latexsym}
\usepackage{amssymb}
\usepackage{amsmath}
\usepackage{dsfont}
\usepackage{url}
\usepackage{graphicx}
\usepackage[applemac]{inputenc}  %mac
\usepackage[english]{babel}
\usepackage{bibentry}

\numberwithin{equation}{section}
\renewcommand{\theequation}{\arabic{section}.\arabic{equation}}

\begin{document}

\title{Effective-one-body Hamiltonian with next-to-leading order spin-spin coupling for two nonprecessing black holes with aligned spins}
\author{Simone Balmelli}
\email{simone.balmelli@gmail.com}
\author{Philippe Jetzer}

\affiliation{Institut f\"ur Theoretische Physik, Universit\"at Z\"urich, Winterthurerstrasse 190, 8057 Z\"urich, Switzerland}
\begin{abstract}
The canonical Arnowitt-Deser-Misner (ADM) Hamiltonian with next-to-leading order (NLO) spin-spin coupling [J. Steinhoff, S. Hergt, and G. Sch\"afer]  is converted into the effective-one-body (EOB) formalism of T. Damour, P. Jaranowski, and G. Sch\"afer for the special case of spinning black hole binaries whose spins are aligned with the angular momentum.
 In particular, we propose to include the new terms by adding a dynamical term of NLO to the Kerr parameter squared entering the effective metric.
 The modified EOB Hamiltonian consistently reduces to the Kerr Hamiltonian as the mass-ratio tends to zero; moreover, it predicts the existence of an innermost stable circular orbit.
 We also derive, for the general case of arbitrarily oriented spins but in the vanishing mass-ratio limit, a coordinate transformation that maps the NLO spin-spin contribution of the ADM Hamiltonian to the EOB Hamiltonian.
\end{abstract}

\pacs{04.25.-g, 04.25.dg}

 \maketitle

\section{Introduction}
\label{sec:intro}

	Coalescing black hole binaries (BHBs) are among the most promising gravitational wave (GW) sources for interferometric, ground-based detectors (like the currently operating LIGO, Virgo and GEO) and the planned space-based detector LISA \cite{amar:12}.
	LIGO and Virgo are going to be upgraded to advanced configurations with a sensitivity improvement of one order of magnitude  \cite{LIGO:10}.
	The volume of space that can be observed will be enlarged by a factor 1000, making a first detection of GW realistic.
	In particular,  for BHBs  with masses of about 10$M_\odot$, a detection rate of roughly 30 events per year seems to be plausible \cite{LIGO:10}. 
	The data analysis needed to extract the GW signal from the background noise is mainly based on the so-called matched-filtering technique, which requires a deep theoretical understanding and a very accurate modeling of the waveforms.
	Since the strongest and most useful signals are emitted in the final stages of the coalescence, a description of the inspiral phase alone (that is already provided with great accuracy by the post-newtonian (PN) theory) is not satisfactory.
	Up to now, the most precise complete waveforms for coalescing BHBs have been generated by numerical relativity simulations.
	However, since the waveforms depend on at least eight parameters (2 for the masses and 6 for the two spins), it is not conceivable to cover the parameter space by a sufficient number of simulations.
	As a consequence, the need has arisen to develop analytical (or semi-analytical) tools to support the results provided by numerical relativity.
	
	Among these methods, the effective-one-body  (EOB)  approach plays a central role.
	Proposed for the first time in 1999 \cite{buon:99}, it is based on the idea of mapping the dynamics of two gravitationally interacting bodies into the geodesics of a fixed, Schwarzschild-like ``effective"  metric, that are usually described by an Hamiltonian.
	The EOB dynamics also includes a dissipative part, that collects the energy and momentum losses of the system and that must be added \textit{ad hoc} into the equations of motion (we refer to \cite{dam:09_EOB, dam:12_EOB} for a review of the EOB formalism).
	EOB models generally involve free parameters that can be calibrated through a comparison with numerical simulations, thus exhibiting a noticeable flexibility.
	Remarkably enough, the first analytical study of the waveform during inspiral and plunge of non spinning binaries has been accomplished within the EOB formalism \cite{buon:00}.
	
	Since then, EOB models have been significantly improved.
	In the non spinning case, Refs.~\cite{dam:07, dam:07_2, dam:08_5, dam:08_4, dam:08_3, dam:09, dam:09_2, pan:09,pan:11,nag:12, dam:13,bern:12} have leaded to increasingly accurate waveforms.
	Relevant analytical improvements have been made especially in the radiation-reaction sector, with the development of a new formalism for the decomposition of multipolar waveforms \cite{dam:09_2, dam:09, pan:11} and, more recently, with the inclusion of the horizon-absorbed GW flux \cite{nag:12, dam:13, bern:12}.
	
	By contrast, waveforms from the coalescence of spinning binaries have not reached a comparable accuracy, in particular for rapidly spinning systems, like extremal BHBs (see e.g. Ref.~\cite{tar:12}).
	This may be due simply to the fact that spin effects beyond the leading-order (LO) of the PN expansion series have been derived only in recent years \cite{fay:06, dam:08_2, por:08, por:10_2, stein:08_ADM_form, stein:08s1s2, por:08_2, por:10, stein:08s^2, levi:10, hart:11_2, hart:11_3, levi:12}, rather than to an intrinsic difficulty of the EOB approach to reproduce the spin interaction.
	Spin effects of coalescing BHBs have been included for the first time into the conservative part of the EOB formalism in Ref.~\cite{dam:01}, according to the natural idea of generalizing the Schwarzschild-like metric into a Kerr-like metric.
	The EOB Hamiltonian proposed there reproduces, when expanded in a PN series, the correct LO spin-orbit and spin-spin couplings.
	Successively, Ref.~\cite{dam:08} extended this model to also reproduce the next-to-leading order (NLO) spin-orbit coupling \cite{dam:08_2}.
	More recently, the next-to-next-to-leading order (NNLO) spin-orbit coupling (derived in Ref.~\cite{hart:11_2}) has been included in the same formalism \cite{nag:11}.
	
	In parallel to this model, a slightly different approach, based on an analytical result reproducing the exact dynamics of a test spin in curved space-time \cite{bar:09}, has been developed in Refs.~\cite{bar:10, bar:11, tar:12}.
	When the test spin limit is not valid, this Hamiltonian reproduces the same spin effects of Refs.~\cite{dam:08,nag:11}, i.e. the LO spin-spin and the NNLO spin-orbit coupling.
	Up to now, this is the only spinning EOB Hamiltonian that has been calibrated to numerical relativity waveforms \cite{pan:10, tar:12}, though only in the case of nonprecessing spins.
	The resulting waveforms are rather accurate, but for nearly extremal black holes (that is, with Kerr parameter $a \gtrsim 0.7M/c^2$) with aligned spin they become unsatisfactory.
	Indeed, compared to the numerical waveforms, they show a dephasing up to 0.8 rad over the entire evolution, while for mildly rotating ( $a \lesssim 0.7M/c^2$) black holes the dephasing does not exceed 0.15 rad \cite{tar:12}.
	
	Among the features of both EOB models, it is worth mentioning the existence of an innermost stable circular orbit (ISCO), which gives a measure of the quantity of GWs emitted before the plunge. 
	As in the case of the exact Kerr metric, the ISCO becomes more bounded for larger, aligned spins, and consequently the GW signal gets stronger.
	In particular, coalescing binaries with aligned, non precessing spins are relevant for GW detection purposes.
 	Indeed, numerical simulations show that BHBs whose spins are aligned with the angular momentum generate a signal 3 times stronger than comparable binaries with spins anti-aligned with respect to the angular momentum and 2 times stronger than comparable non spinning binaries.
	The observational volume is thus 27 times and 8 times larger, respectively \cite{reiss:09}.
	 Moreover, the alignment between spins and angular momentum seems to be favoured by accretion mechanisms in gas-rich environments \cite{bog:07}.

	NLO spin-spin effects have already been calculated a few years ago \cite{por:08, por:10_2, stein:08s1s2, por:08_2, por:10, stein:08s^2, levi:10}.
	Motivated by the above arguments, this paper attempts to improve the EOB Hamiltonian of Refs.~\cite{dam:01,dam:08,nag:11} by including the NLO spin-spin coupling in the special case of  BHBs whose spins are aligned (or anti-aligned) with the angular momentum.
	More precisely, we show that it is possible to reproduce the correct NLO spin-spin terms by adding a dynamical NLO term to the square of the Kerr parameter of the effective metric.
	The price to pay is that an effective spin depending on the dynamical variables may introduce physical inconsistencies like the violation of the Kerr bound.
	However, we show that this can be avoided by the appropriate introduction of an additional NNLO term to the effective squared spin .
	Furthermore, the old (variable-independent) effective spin is recovered in the small mass-ratio limit, as required by consistency. 
	Finally, we show that the existence of an ISCO is preserved.
	
	The paper is structured as follows: in Sec.~\ref{sec:ADM} we present the PN expanded Hamiltonian provided by the ADM theory, and simplify the NLO spin-spin Hamiltonian of Refs.~\cite{ stein:08s1s2, stein:08s^2} using the center of mass coordinates and taking into account the alignment constraint.
	In Sec.~\ref{sec:transf} we discuss the mapping between the ADM and the EOB dynamics, performing the appropriate canonical transformations in the case of rapidly rotating spins. 
	An additional canonical transformation which is quadratic in the spins and of NLO accuracy is introduced.
	In Sec.~\ref{sec:PN_EOB} we summarize the structure of the EOB Hamiltonian as given by Ref.~\cite{dam:08} and calculate the corresponding 3PN spin-spin contribution.
	Sec.~\ref{sec:NLO_inc} completes the matching between the ADM and the EOB dynamics, proposing a modification of the spin parameter entering the effective Kerr-like metric in order to reproduce the desired spin-spin coupling.
	For the general case of arbitrarily oriented spins, we derive the canonical transformation that is needed for ensuring the reduction of a future, complete EOB Hamiltonian with NLO spin-spin coupling to the Kerr Hamiltonian whenever the mass-ratio tends to zero.
	In Sec.~\ref{sec:disc} we show that the modified Hamiltonian still predicts the existence of an ISCO, and discuss the problems arising from the dependency of the modified effective squared spin on the dynamical variables.	
	Finally, in the Appendix we show that the alignment between spins and total angular momentum is conserved during the dynamical evolution at least at the PN order we are dealing with.

\section{Next-to-leading order, spin-spin Hamiltonian in ADM coordinates}
\label{sec:ADM}

	The PN-expanded ADM Hamiltonian for two gravitationally interacting and spinning point masses can be decomposed as

		\begin{alignat}{1}	
		\label{eq:ADM_Ham}
			H\left(\bm{x},\bm{p},\bm{S}_1,\bm{S}_2\right) = &\,\,H_{\text{o}}\left(\bm{x},\bm{p}\right)+H_{\text{so}}\left(\bm{x},\bm{p},\bm{S}_1,\bm{S}_2\right)\nonumber\\
			& + H_{\text{ss}}\left(\bm{x},\bm{p},\bm{S}_1,\bm{S}_2\right)\nonumber\\				& + ...,
		\end{alignat}
	where $H_{\text{o}}$ denotes the purely orbital part, while $H_{\text{so}}$ and $H_{\text{ss}}$ describe the spin-orbit and the spin-spin interaction, respectively.
	
	It may be convenient to introduce the center of mass frame ($\bm{R}\equiv \bm{x}_1- \bm{x}_2$, $\bm{P} \equiv \bm{p}_1-\bm{p}_2$) and the corresponding rescaled coordinates $\bm{r} \equiv \bm{R}/M$, $\bm{p} \equiv \bm{P}/\mu$, where  $M \equiv m_1+m_2$ is the total mass and $\mu \equiv m_1\,m_2/M$ the reduced mass.
	Moreover, we define the symmetric mass-ratio $\nu \equiv \mu/M$ and use the notation $r \equiv |\bm{r}|$, $\bm{n} \equiv \bm{r}/r$.
	The spins can be rescaled according to $\hat{\bm{S}}_a \equiv \bm{S}_a/(M \mu)$ (but, as discussed below, we will use a different notation).
	Finally, we rescale the Hamiltonian according to $\hat{H} \equiv H/\mu$ \cite{dam:08, nag:11}.
	For simplicity, we use units with $G \equiv 1$. 
	The PN structure of the orbital Hamiltonian is

		\begin{alignat}{1}
			\hat{H}_{\text{o}}= &\,\, \frac{c^2}{\nu} +\hat{H}_{\text{o}}^\text{N}+\hat{H}_{\text{o}}^\text{1PN}+\hat{H}_{\text{o}}^\text{2PN}\nonumber\\
			&+\hat{H}_{\text{o}}^\text{3PN}+ \mathcal{O}\left(\frac{1}{c^8}\right).
		\end{alignat}	
	The Newtonian term is simply
	
		\begin{equation}
		\label{eq:N_Ham}
			\hat{H}_{\text{o}}^\text{N}= \frac{\bm{p}^2}{2}-\frac{1}{r},
		\end{equation}
	while the 1PN one reads

		\begin{equation}
			\hat{H}_{\text{o}}^\text{1PN}= \frac{1}{c^2}\bigg[\frac{(3\nu-1)}{8}\bm{p}^4 - \frac{(3+\nu)}{2}\frac{\bm{p}^2}{r} - \frac{\nu}{2}\frac{(\bm{n} \cdot \bm{p})^2}{r}  +\frac{1}{2 r^2}\bigg].
		\end{equation}	
		
	For an explicit expression of the 2PN accurate Hamiltonian see Ref.~\cite{dam:88}, and for the 3PN accurate one Ref.~\cite{jar:98}. 	
	The expansion of the spin-dependent part can be written as
	
		\begin{alignat}{2}
			H_{\text{so}}&= \,\,&& H_{\text{so}}^{\text{LO}}+H_{\text{so}}^{\text{NLO}} +...\nonumber\\	
				& = && \sum_{a\equiv1,2} \bm{S}_a \cdot \left(\bm{\Omega}_{a}^{\text{LO}} +\bm{\Omega}_{a}^{\text{NLO}}+...  \right)\nonumber\\
			H_{\text{ss}}&= && \left(H_{\text{S}_1^2}^{\text{LO}}+H_{\text{S}_2^2}^{\text{LO}}+H_{\text{S}_1\text{S}_2}^{\text{LO}}\right)\nonumber\\
				& && +\left(H_{\text{S}_1^2}^{\text{NLO}}+H_{\text{S}_2^2}^{\text{NLO}} + H_{\text{S}_1\text{S}_2}^{\text{NLO}}\right)\nonumber\\
				& && + ...\nonumber\\
				& \equiv && H_{\text{ss}}^{\text{LO}}+ H_{\text{ss}}^{\text{NLO}} + ...
		 \end{alignat}		
	The terms composing $H_{\text{so}}$ have been derived in Ref.~\cite{dam:08_2} (up to NLO) and in Ref.~\cite{hart:11_2} (at NNLO).
	
	Both $H_{\text{so}}$ and $H_{\text{ss}}$ formally start at 1PN ($\propto \,1/c^2$).
	However, the real PN order depends on the order of magnitude of the spins.
	For example, for extremal black holes the spins  are proportional to $ 1/c$ , which corresponds to 0.5PN.
	This implies that the leading order term $H_{\text{so}}^{\text{LO}}$ of the spin-orbit Hamiltonian is 1.5PN accurate, while  $H_{\text{ss}}^{\text{LO}}$ is 2PN accurate.
	In this paper, we rescale the spins in such a way that the powers of $c^{-1}$ label the true PN order in the case of extremal black holes.
	We write
	
		\begin{equation}
			\bm{S}_a = \frac{m_a^2}{c} \bm{\chi}_a,		
		\end{equation} 
	where $|\bm{\chi_a}| \leq 1$ is dimensionless. 
		
	The leading-order spin-spin contribution can be written as \cite{dam:01}
	
		\begin{equation}
		\label{eq:ADMLOss}
			\hat{H}_{\text{ss}}^{\text{LO}} = \frac{1}{2c^4}\frac{3(\bm{n} \cdot \bm{\chi}_0)^2-(\bm{\chi}_0)^2}{r^3},
		\end{equation}
	
	with the linear combination
		
		\begin{equation}
		\label{eq:S_0}
			\bm{\chi}_0 = \frac{m_1}{M}\bm{\chi}_1+  \frac{m_2}{M}\bm{\chi}_2.
		\end{equation}
	Finally, the next-to-leading order, spin-squared Hamiltonian $H_{\text{S}_1^2}^{\text{NLO}}$ has been derived explicitly in Ref.~\cite{stein:08s^2}, and the spin(1)-spin(2) Hamiltonian $H_{\text{S}_1 \text{S}_2}^{\text{NLO}}$ in Ref.~\cite{stein:08s1s2}.
	After going to the rescaled center of mass coordinates (according to the above prescriptions), they read
	
	\begin{widetext}
		\begin{subequations}	
			\begin{alignat}{1}
			\label{eq:NLOS_{1}^2}
				\hat{H}_{\text{S}_1^2}^{\text{NLO}} = & \quad \frac{\nu}{c^6\, r^3}\frac{m_1}{m_2} \bigg[  + \frac{1}{4} \left(  1-2\nu-\frac{m_1}{m_2}\nu \right)(\bm{\chi}_1 \cdot \bm{p})^2 + \frac{3}{8} \left(  1-4\nu-\frac{m_1}{m_2}\nu\right) \bm{\chi}_1^2( \bm{p} \cdot \bm{n})^2\nonumber\\			
					& + \frac{3}{8} \left( - 1+8\nu + 7\frac{m_1}{m_2}\nu\right)( \bm{\chi}_1 \cdot \bm{n})^2 \bm{p}^2 + \frac{3}{4} \left(-1 +\frac{m_1}{m_2}\nu  \right)( \bm{\chi}_1 \cdot \bm{p} ) ( \bm{\chi}_1 \cdot \bm{n})( \bm{p} \cdot \bm{n}) \nonumber\\	
					& + \frac{15}{4} \nu ( \bm{\chi}_1 \cdot \bm{n})^2( \bm{p} \cdot \bm{n})^2 - \frac{3}{4}\nu \left(1+ \frac{m_1}{m_2} \right) \bm{\chi}_1^2 \bm{p}^2 \bigg]\nonumber\\
				& + \frac{\nu^2}{c^6\,r^4} \left(1 +  \frac{m_1}{m_2} \right) \bigg[\left(  3+\frac{5}{2}\frac{m_1}{m_2}\right)\bm{\chi}_1^2 - \left( 7+\frac{9}{2}\frac{m_1}{m_2} \right)(\bm{\chi}_1\cdot \bm{n})^2 \bigg],
			\end{alignat}	
					
			\begin{alignat}{1}
			\label{eq:NLOS1S2}
				\hat{H}_{\text{S}_1\text{S}_2}^{\text{NLO}} = &  \quad \frac{3\nu^2}{2c^6\,r^3} \bigg [-\left( \frac{5}{2} +\frac{m_2}{m_1} + \frac{m_1}{m_2}\right)\big((\bm{p}\wedge\bm{\chi}_1)\cdot \bm{n}\big)\big((\bm{p}\wedge\bm{\chi}_2)\cdot \bm{n}\big)  + 5(\bm{\chi}_1 \cdot \bm{n})(\bm{\chi}_2 \cdot \bm{n})(\bm{p} \cdot \bm{n})^2\nonumber\\
					& + (\bm{\chi}_1 \cdot \bm{n})(\bm{\chi}_2 \cdot \bm{n})\bm{p}^2 - \left(2 + \frac{m_1}{m_2}\right)(\bm{\chi}_1 \cdot \bm{p})(\bm{\chi}_2 \cdot \bm{n})(\bm{p} \cdot \bm{n})\nonumber\\
					&- \left(2 + \frac{m_2}{m_1}\right)(\bm{\chi}_1 \cdot \bm{n})(\bm{\chi}_2 \cdot \bm{p})(\bm{p} \cdot \bm{n}) - \frac{1}{6}(\bm{\chi}_1 \cdot \bm{p})(\bm{\chi}_2 \cdot \bm{p})\nonumber\\
					& -\frac{1}{6}(\bm{\chi}_1 \cdot \bm{\chi}_2)\bm{p}^2 +\left(1+ \frac{m_2}{m_1}+ \frac{m_1}{m_2} \right)(\bm{\chi}_1 \cdot \bm{\chi}_2)(\bm{p}\cdot \bm{n})^2 \bigg]\nonumber\\
				& + \frac{6\nu}{c^6\, \,r^4} \bigg[ \, (\bm{\chi}_1 \cdot \bm{\chi}_2) - 2(\bm{\chi}_1 \cdot \bm{n})(\bm{\chi}_2 \cdot \bm{n})\bigg].
			\end{alignat}
		\end{subequations}
	\end{widetext}	
	
	The Hamiltonian $H_{\text{S}_2^2}^{\text{NLO}}$ can be obtained from Eq.~(\ref{eq:NLOS_{1}^2}) by simply exchanging the particle labels 1 and 2.
	
	If we are interested in the special case where the angular momentum is aligned with both spins, we can set $(\bm{\chi}_a \cdot \bm{p}) = (\bm{\chi}_a \cdot \bm{n})=0$.
	Then, the sum of the above Hamiltonians reduces to
	
	\begin{widetext}
		\begin{alignat}{1}
		\label{eq:NLOss:scal}
			\hat{H}_{\text{ss,aligned}}^{\text{NLO}} =   \frac{1}{c^6}\frac{\nu}{8\, r^3}\Bigg\{&\frac{m_1}{m_2}\bigg[-6\nu\left(1+\frac{m_1}{m_2} \right)\bm{p}^2+\left(3-12\nu-3\nu\frac{m_1}{m_2}\right)(\bm{n}\cdot\bm{p})^2+\left(24-4\nu-4\nu\frac{m_1}{m_2}\right)\frac{1}{r}\bigg]\bm{\chi}_1^2\nonumber\\
				+& \frac{m_2}{m_1}\bigg[-6\nu\left(1+\frac{m_2}{m_1} \right)\bm{p}^2+\left(3-12\nu-3\nu\frac{m_2}{m_1}\right)(\bm{n}\cdot\bm{p})^2+\left(24-4\nu-4\nu\frac{m_2}{m_1}\right)\frac{1}{r}\bigg]\bm{\chi}_2^2\nonumber\\
				+&\bigg[-\nu \left(32 +12\left(\frac{m_1}{m_2}+\frac{m_2}{m_1}\right)\right)\bm{p}^2+\nu\left(42+ 24\left(\frac{m_1}{m_2}+\frac{m_2}{m_1}\right)\right)(\bm{n}\cdot\bm{p})^2 + \frac{6}{r}\bigg](\bm{\chi}_1\cdot \bm{\chi}_2)\Bigg\}.
		\end{alignat}
	\end{widetext}	
		
\section{Transformation from ADM to EOB coordinates}
\label{sec:transf}

	In order to translate the ADM Hamiltonians into the EOB formalism, some appropriate coordinate transformations have to be performed.
	A first step is the purely orbital canonical transformation $G_o\left(\bm{r},\bm{p}\right)$ \cite{buon:99}.
	Moreover, since this work should be consistent with previous ones \cite{dam:01, dam:08, nag:11}, we also have to take into account the canonical transformations that have been applied there.
	
	A generating function $G\left(\bm{r},\bm{p}' \right)$ transforms the coordinates according to
		\begin{subequations}	
		\label{eq:coord_transf}
			\begin{alignat}{1}
				\bm{r}' &= \bm{r} + \{\bm{r}, G\left(\bm{r},\bm{p}' \right)\}\\
				\bm{p}' &= \bm{p} + \{\bm{p}', G\left(\bm{r},\bm{p}' \right)\},
			\end{alignat}
		\end{subequations}
	where the derivatives inside the Poisson Brackets are taken with respect to $\bm{r}$ and $\bm{p}'$.
	The time independence of $G$ ensures that the transformed Hamiltonian is numerically invariant, i.e. $H'\left(\bm{q}' \right)=H\left(\bm{q} \right)$.
	Provided that $G$ can be treated as a small, perturbative factor, one can obtain the transformed Hamiltonian $H'$ by inserting Eq.~(\ref{eq:coord_transf}) into the numerical invariance condition.
	At linear order in $G$ one has then
	
	 	\begin{equation}
			H'\left(\bm{q}'\right) = H\left(\bm{q}'\right)+  \{G\left( \bm{q}'\right),H\left(\bm{q}'\right)\}.
		\end{equation}
	
	The 1PN orbital generating function 
	
		\begin{equation}
		\label{eq:1_gen_func}
			\hat{G}_{\text{o}}^{\text{1PN}} =\frac{1}{c^2}(\bm{r} \cdot \bm{p}')\left( -\frac{\nu}{2} \bm{p}'^2 + \left( 1 + \frac{\nu}{2}\right)\frac{1}{r} \right)
		\end{equation}
	transforms $\hat{H}_\text{ss}^\text{NLO}$ according to
	
		\begin{alignat}{1}
		\label{eq:1_can_trans}
			\hat{H}_{\text{ss}}^{\text{NLO}\prime}\left(\bm{r}',\bm{p}',\bm{\chi}_1,\bm{\chi}_2\right) = &\hat{H}_{\text{ss}}^{\text{NLO}}\left(\bm{r}',\bm{p}',\bm{\chi}_1,\bm{\chi}_2\right)\nonumber\\
				 &+ \{\hat{G}_{\text{o}}^{\text{1PN}},\hat{H}_{\text{ss}}^{\text{LO}}\}\left(\bm{r}',\bm{p}',\bm{\chi}_1,\bm{\chi}_2\right).
		\end{alignat}
	As a second step, we need the canonical transformation that has been employed to obtain the LO, spin-spin Hamiltonian in EOB coordinates \cite{dam:01}.
	As pointed out in Ref.~\cite{bar:11}, the corresponding generating function is given by
	
		\begin{alignat}{1}
		\label{eq:2_gen_func}
			\hat{G}_{\text{ss}}^\text{LO}\left(\bm{r},\bm{p}',\bm{\chi}_1,\bm{\chi}_2\right) &= - \frac{1}{c^4}\frac{1}{2\,r^2}\Big \{ \big [ \bm{\chi}_0^2- (\bm{\chi}_0 \cdot \bm{n})^2 \big ] (\bm{r} \cdot \bm{p}')\nonumber\\
				& +  \big(\bm{\chi}_0 \cdot \bm{n}\big)\left(\bm{r} \times \bm{p}'\right)\cdot \big(\bm{\chi}_0 \times \bm{n}\big) \Big \},
		\end{alignat}
	where $\chi_0$ has already been defined in Eq.~(\ref{eq:S_0}).
	When applied onto the transformed orbital Hamiltonian 
	
		\begin{equation}
 			\hat{H}_{\text{o}}^{\text{1PN} \prime} =  \hat{H}_{\text{o}}^{\text{1PN}} +  \{ \hat{G}_{\text{o}}^{\text{1PN}}, \hat{H}_{\text{o}}^{\text{N}} \big \},
		\end{equation}
	$\hat{G}_{\text{ss}}^\text{LO}$ gives rise to some additional NLO, spin-spin terms:
	
		\begin{alignat}{1}
			\hat{H}_{\text{ss}}^{\text{NLO}\prime \prime}\left(\bm{r}'',\bm{p}'',\bm{\chi}_1,\bm{\chi}_2\right) = &\hat{H}_{\text{ss}}^{\text{NLO}\prime}\left(\bm{r}'',\bm{p}'',\bm{\chi}_1,\bm{\chi}_2\right)\nonumber\\
				 &+ \{\hat{G}_{\text{ss}}^{\text{LO}},\hat{H}_{\text{o}}^{\text{1PN}\prime}\}\left(\bm{r}'',\bm{p}'',\bm{\chi}_1,\bm{\chi}_2\right).
		\end{alignat}
	Now, we perform an additional NLO coordinate transformation $\hat{G}_{\text{ss}}^{\text{NLO}}$ which is quadratic in the spins.
	The Hamiltonian is transformed according to
		
		\begin{alignat}{1}
		\label{eq:3_can_transf}
			\hat{H}_{\text{ss}}^{\text{NLO}\prime \prime \prime}\left(\bm{r}''',\bm{p}''',\bm{\chi}_1,\bm{\chi}_2\right) = &\hat{H}_{\text{ss}}^{\text{NLO}\prime \prime}\left(\bm{r}''',\bm{p}''',\bm{\chi}_1,\bm{\chi}_2\right)\nonumber\\
				 &+ \{\hat{G}_{\text{ss}}^{\text{NLO}},\hat{H}_{\text{o}}^{\text{N}}\}\left(\bm{r}''',\bm{p}''',\bm{\chi}_1,\bm{\chi}_2\right).
		\end{alignat}
	Notice that the orbit, spin-orbit and LO spin-spin terms are not affected by this transformation, which can thus be safely used without compromising the results obtained by the previous papers.
	When taking into account the alignment constraint $(\bm{\chi}_a \cdot \bm{p}) = (\bm{\chi}_a \cdot \bm{n)} = 0$, the most general form of $\hat{G}_{\text{ss}}^{\text{NLO}}$ is simply

		\begin{alignat}{1}
 			\hat{G}_{\text{ss,al}}^{\text{NLO}} =& \frac{1}{c^6\, r}\bigg\{  \bigg [  \alpha_{11} \, \bm{p}^2(\bm{n}\cdot \bm{p}) + \beta_{11} (\bm{n}\cdot \bm{p})^3  + \gamma_{11} \frac{(\bm{n}\cdot \bm{p})}{r}\bigg]  \bm{\chi}_1^2\nonumber\\
				+ & \bigg [ \alpha_{12} \, \bm{p}^2(\bm{n}\cdot \bm{p}) + \beta_{12} (\bm{n}\cdot \bm{p})^3 + \gamma_{12} \frac{(\bm{n}\cdot \bm{p})}{r}\bigg]  (\bm{\chi}_1\cdot \bm{\chi}_2)\nonumber\\
				+ &  \bigg [ \alpha_{22} \, \bm{p}^2(\bm{n}\cdot \bm{p}) + \beta_{22} (\bm{n}\cdot \bm{p})^3 + \gamma_{22} \frac{(\bm{n}\cdot \bm{p})}{r}\bigg]  \bm{\chi}_2^2\bigg\},
		\end{alignat}
	where $\alpha_{ab}$, $\beta_{ab}$ and $\gamma_{ab}$ are gauge parameters. 
	
	This third transformation is the last step necessary to translate the ADM formalism into the EOB one.
	The EOB model correctly reproduces the NLO spin-spin effects if $\hat{H}_{\text{ss}}^{\text{NLO}\prime \prime \prime}$ is formally equal to the corresponding contribution $\hat{H}_\text{EOB,ss}^\text{NLO}$ from the PN expansion of the ``real'' EOB Hamiltonian $\hat{H}_{\text{EOB}}$ (see Sec.~\ref{sec:PN_EOB}).
	In order to simplify the notation, we omit the triple prime so that $\bm{r}$, $\bm{p}$ now denote the new (rescaled) EOB coordinates appearing in Eq.~(\ref{eq:3_can_transf}).
	This notation will be adopted until the end of the paper.
	
\section{PN expansion of the EOB Hamiltonian with leading-order spin-spin coupling}
\label{sec:PN_EOB}
	
	We remember that this paper closely follows the lineage of Refs.~\cite{dam:01, dam:08, nag:11}.
	We shortly review the basic structure of the formalism that has been employed there.
	The EOB Hamiltonian takes the form
	
		\begin{equation}
		\label{eq:HEOB}
			H_{\text{EOB}} = M\,c^2 \sqrt{1+ 2\nu\left( \frac{H_{\text{eff}}}{\mu \, c^2}-1\right)},
		\end{equation}
	where
	
		\begin{equation}
		\label{eq:Heff}
			H_{\text{eff}} = H_{\text{eff,0}} + \Delta H_{\text{eff,so}} 
		\end{equation}
	is the so-called effective Hamiltonian. 
	The term $\Delta H_{\text{eff,so}}$ (see Eq.~(4.16) of Ref.~\cite{dam:08}) has been introduced in order to correctly reproduce spin-orbit interaction up to NNLO.
	It has a linear dependence on the ``test-spin''
	
		\begin{equation}
		\label{eq:sigma}
			\bm{\sigma} = \frac{1}{2}\left( g_S^\text{eff} \bm{S} + g_{S^*}^\text{eff} \bm{S}^*\right)-\bm{S}_0,
		\end{equation}
	that is defined through the linear combinations of spins $\bm{S} \equiv \bm{S}_1 +\bm{S}_2$ and $\bm{S}^* \equiv  (m_2/m_1)\bm{S}_1 + (m_1/m_2)\bm{S_2}$, and through the ``gyro-gravitomagnetic" ratios $g_S^\text{eff}$ and $g_{S^*}^\text{eff}$ \cite{dam:08,nag:11}.	
	
	The Hamiltonian $H_{\text{eff,0}}$ describes the motion of a test particle of mass $\mu$ in an external metric $g_{\text{eff}}$. It can be written as
	
		\begin{equation}
			H_{\text{eff,0}}=  \beta^{i} P_{i}\, c + \alpha \, c \sqrt{\mu^2\, c^2 + \gamma^{ij}\,P_i\,P_j +Q_4(P_i)},  
		\end{equation}	  
	where $Q_4(P_i)$ is a quartic-in-momenta term \cite{dam:00, dam:01} and where $\alpha$, $\beta^i$ and $\gamma^{ij}$ are the lapse, shift and 3-metric of $g_{\text{eff}}$, i.e.
	
		\begin{subequations}	
			\begin{alignat}{1}
				\alpha &= \frac{1}{\sqrt{- g_{\text{eff}}^{00}}} \\
				\beta^i &= \frac{g_{\text{eff}}^{0i}}{g_{\text{eff}}^{00}}\\
				\gamma^{ij} &= g_{\text{eff}}^{ij}-  \frac{g_{\text{eff}}^{0i}\,g_{\text{eff}}^{0j}}{g_{\text{eff}}^{00}}.
			\end{alignat}	
		\end{subequations}
	The metric $g_{\text{eff}}$ is a $\nu$-deformed Kerr metric for a central mass $M$ and effective Kerr parameter 
	
		\begin{equation}
			a_0= \bigg |\left(1+\frac{m_2}{m_1}\right)\frac{\bm{S}_1}{M\,c}+\left(1+\frac{m_1}{m_2}\right)\frac{\bm{S}_2}{M\,c}\,\bigg|.
		\end{equation}

	In Ref.~\cite{dam:01}, $g_{\text{eff}}$ was first written in Boyer-Linquist-like coordinates and successively transformed into Cartesian-like coordinates, in order to allow the spins to rotate in any direction.
	Since we will finally keep the spins fixed along the $\bm{e}_3$ axis, however, a Boyer-Lindquist-like coordinate system is appropriate for our purposes. 
	Using the index notation $0 = t$, $i = R,\,\theta,\,\varphi$ and with the additional notation
	
		\begin{equation}
			\rho = \sqrt{R^2 + a_0^2 \cos^2(\theta)}, 
		\end{equation}	
	the effective metric reads

		\begin{subequations}	
		\label{eq:metric}
			\begin{alignat}{1}
				g_{\text{eff}}^{tt} &= \frac{1}{\rho^2} \left( a_0^2 \sin^2(\theta) - \frac{\left(a_0^2 +R^2\right)^2}{\Delta_t(R)}\right)\\
				g_{\text{eff}}^{RR} &= \frac{\Delta_R(R)}{\rho^2}\\
				g_{\text{eff}}^{\theta \theta} &= \frac{1}{\rho^2}\\
			 	g_{\text{eff}}^{\varphi \varphi } &=\frac{1}{\rho^2}\left( \frac{1}{\sin^2(\theta)}- \frac{a_0^2}{\Delta_t(R)}\right)\\
			\label{eq:g^{tphi}}
				g_{\text{eff}}^{t \varphi } &=\frac{a_0}{\rho^2}  \left( 1- \frac{a_0^2+R^2}{\Delta_t(R)}\right).
			\end{alignat}
		\end{subequations}
	The functions $\Delta_t$ and $\Delta_R$ encode, according to the EOB philosophy, the (Padé-resummed) PN terms in a $\nu$-dependent way.
	They are defined through

		\begin{subequations}
			\begin{alignat}{1}
				\Delta_t &= R^2 P^n_m\bigg[ A(u) + u^2 \frac{c^4\,a_0^2}{M^2}\bigg]\\
				\label{eq:D^-1}
				\Delta_R &= \Delta_t \,D^{-1},
			\end{alignat}
		\end{subequations}
	where $P^n_m$ denotes the action of taking the (n,m)-Pad\'e approximant with respect to the variable $u = M /(c^2\, R)$.
	Finally, $A$ and $D^{-1}$ are Schwarzschild-like metric coefficients, which at 3PN accuracy are given by

		\begin{subequations}
		\label{eq:AD^-1}
			\begin{alignat}{1}
				A(u) &= 1 -2u + 2\nu\,u^3 + \left(\frac{94}{3} - \frac{41}{32}\pi^2\right)\nu\,u^4\\
				D^{-1}(u) &= 1 + 6\nu\,u^2 + 2(26-3\nu)\nu\,u^3.
			\end{alignat}
		\end{subequations}
	For $\nu = 0$ both $\Delta_t$ and $\Delta_R$ reduce to $\Delta = R^2 -2MR/c^2+a_0^2$, so that the exact Kerr metric is recovered.

	When expanded in PN orders, the elements of $ g_{\text{eff}}$ form a series in powers of $M/(c^2R)$.
	The expansion has to be performed with respect to the Kerr parameter $a_0$ too.
	This is done writing 
	 
	 	\begin{equation}
	 		a_0 \equiv \frac{M}{c^2} \chi_0.
		\end{equation}
	Notice that $\chi_0$ is defined consistently with respect to Eq.~(\ref{eq:S_0}).
	For completeness, we write the expansion of the lapse, shift and 3-metric up to 3PN:
	
		\begin{subequations}
			\begin{alignat}{2}
				\alpha &= && 1 - \frac{M}{c^2 R} - \frac{1}{2} \left(\frac{M}{c^2 R}\right)^2\nonumber\\
					& && + \left((\bm{n} \cdot \bm{\chi}_0)^2 -\frac{1}{2}+\nu \right) \left( \frac{M}{c^2 R}\right)^3 + \mathcal{O}\left( \frac{1}{c^8}\right)\\
				\beta^\varphi &= &&\frac{2 \chi_0}{R} \left(\frac{M}{c^2 R}\right)^2 + \mathcal{O}\left( \frac{1}{c^8}\right)\\
				\gamma^{RR} &= &&  1 - 2 \left(\frac{M}{c^2 R}\right) + \left( 6\nu + \bm{\chi}_0^2 -(\bm{n} \cdot \bm{\chi}_0)^2 \right) \left(\frac{M}{c^2 R}\right)^2 \nonumber\\
					& && + \left( 42 \nu - 6 \nu^2 +2(n\bm{\chi}_0)^2 \right) \left(\frac{M}{c^2 R}\right)^3 +\mathcal{O}\left( \frac{1}{c^8}\right)\\
				\gamma^{\theta \theta} & = && \frac{1}{R^2}\bigg [1 -(\bm{n} \cdot \bm{\chi}_0)^2\left(\frac{M}{c^2 R}\right)^2 + \mathcal{O}\left( \frac{1}{c^8}\right) \bigg ]\\
				\gamma^{\varphi \varphi} & = && \frac{1}{\sin^2(\theta)\, R^2} \bigg[1- \bm{\chi}_0^2\left(\frac{M}{c^2 R}\right)^2\nonumber\\
					& && - 2\left( \bm{\chi}_0^2 -(\bm{n} \cdot \bm{\chi}_0)^2\right) \left(\frac{M}{c^2 R}\right)^3+ \mathcal{O}\left( \frac{1}{c^8}\right)  \bigg ].
			\end{alignat}
		\end{subequations}
	We do not write explicitly the whole, straightforward expansion of $H_\text{eff}$ up to 3PN, but just the spin-spin terms.
	At first we redefine the variables, introducing a notation compatible with the calculations of Sec.~\ref{sec:transf}:	
	
		\begin{subequations}	
			\begin{alignat}{1}
			 	\bm{p}^2 &\equiv \frac{1}{\mu^2}\left(P_R^2 + \frac{P_\theta^2}{R^2}+ \frac{P_\varphi^2}{R^2\sin^2(\theta)}\right) \\
				\left(\bm{n}\cdot \bm{p} \right) &\equiv  \frac{P_R}{\mu}\\
				r &\equiv \frac{R}{M}.
			\end{alignat}
		\end{subequations}
	We then have
	
		\begin{alignat}{1}
		\label{eq:H_{eff,ss}^{LO}}
			\hat{H}_{\text{eff,ss}}^{\text{LO}} = &\; \frac{1}{c^4}\bigg [\left( \frac{(\bm{n}\cdot \bm{p})^2}{r^2} - \frac{1}{2} \frac{\bm{p}^2}{r^2} \right)\bm{\chi}_0^2 +\frac{1}{2} \frac{\left( \bm{p} \cdot \bm{\chi}_0\right)^2}{r^2}\nonumber\\
				&- \frac{\left( \bm{n} \cdot \bm{p}\right)\left( \bm{p} \cdot \bm{\chi}_0\right) \left( \bm{n} \cdot \bm{\chi}_0\right)}{r^2} + \frac{ \left( \bm{n} \cdot \bm{\chi}_0\right)^2}{r^3} \bigg]
		\end{alignat}	
		
		\begin{alignat}{1}
		\label{eq:H_{eff,ss}^{3PN}}
			\hat{H}_{\text{eff,ss}}^{\text{NLO}} =&  \frac{1}{c^6}\bigg [\left( \frac{1}{4} \frac{\bm{p}^4}{r^2} - \frac{1}{2}\frac{(\bm{n}\cdot \bm{p})^2\bm{p}^2}{r^2}- \frac{1}{2}\frac{\bm{p}^2}{r^3}+ \frac{2}{r^4}\right)\bm{\chi}_0^2 \nonumber\\
				&+ \left(  \frac{3}{2}\frac{\bm{p}^2}{r^3}+ \frac{\left(\bm{n}\cdot \bm{p}\right)^2}{r^3}- \frac{1}{r^4}\right) (\bm{n} \cdot \bm{\chi}_0)^2\nonumber\\
				&+ \left( -\frac{1}{4} \frac{\bm{p}^2}{r^2} + \frac{1}{2\, r^3} \right)\left( \bm{p} \cdot \bm{\chi}_0\right)^2\nonumber\\
				&+ \left( \frac{1}{2} \frac{\bm{p}^2\left( \bm{n} \cdot \bm{p}\right)}{r^2}- \frac{\left( \bm{n} \cdot \bm{p}\right)}{r^3}\right)\left(\bm{p} \cdot \bm{\chi}_0\right)\left(\bm{n} \cdot \bm{\chi}_0\right) \bigg ].
		\end{alignat}
	It is worth mentioning that the (rescaled) spin-spin contributions turn out to be independent of the deformation parameter $\nu$, and can therefore be directly compared with the PN expanded Kerr Hamiltonian.
	Actually, one can check that Eq.~(\ref{eq:H_{eff,ss}^{LO}}) corresponds to Eq.~(5.55) of Ref.~\cite{bar:11}.

	Finally, the ``effective'' dynamics has to be mapped, according to Eq.~(\ref{eq:HEOB}), to the ``real'' dynamics described by $\hat{H}_\text{EOB}$.
	From the inverse relation
	
		\begin{equation}
			\hat{H}_\text{eff}= \frac{\mu^2\, c^2\,\hat{H}_\text{EOB}^2-m_1^2\,c^4-m_2^2\,c^4}{2\,m_1\,m_2\,c^4}
		\end{equation}
	it is easily found that
	
		\begin{equation}
			 \hat{H}_{\text{eff,ss}}^{\text{NLO}} = \hat{H}_{\text{EOB,ss}}^{\text{NLO}}+ \frac{\nu}{c^2} \hat{H}_{\text{EOB,o}}^\text{N}\hat{H}_{\text{EOB,ss}}^{\text{LO}},
		\end{equation}	
	where $\hat{H}_{\text{EOB,o}}^\text{N}$ and $\hat{H}_{\text{EOB,ss}}^{\text{LO}}$ are left unmodified by the above mapping and can thus be obtained directly from the PN expansion of $\hat{H}_\text{eff}$.
	The first one is simply the Newtonian Hamiltonian
	
		\begin{equation}
			\hat{H}_{\text{EOB,o}}^\text{N}= \frac{\bm{p}^2}{2}-\frac{1}{r},
		\end{equation}
	while $\hat{H}_{\text{EOB,ss}}^{\text{LO}}$ is given by Eq.~(\ref{eq:H_{eff,ss}^{LO}}).
	As a consistence check of the mapping between ADM and EOB coordinates, $\hat{H}_{\text{EOB,ss}}^{\text{LO}}$ can also be obtained by adding to (\ref{eq:ADMLOss}) the Poisson Bracket formed by the terms given in Eqs.~(\ref{eq:2_gen_func}) and (\ref{eq:N_Ham}).
	
\section{Including next-to-leading order spin-spin effects for equatorial orbits and aligned spins}
\label{sec:NLO_inc}	
	Let us denote the EOB Hamiltonian of Sec.~\ref{sec:PN_EOB} with an additional label ``old'', stressing the fact that we are now searching a new Hamiltonian $\hat{H}_\text{EOB}$ that correctly reproduces the NLO spin-spin terms.	
	The correspondence between  ADM and EOB coordinates that has been worked out in Sec.~\ref{sec:transf} requires that $\hat{H}_{\text{EOB,ss}}^{\text{NLO}}$ must be equal to $\hat{H}_{\text{ss}}^{\text{NLO} \prime \prime \prime}$ (\ref{eq:3_can_transf}).
	Writing 
	
		\begin{equation}
			\hat{H}_\text{EOB,ss}^\text{NLO} = \hat{H}_\text{EOB,ss}^\text{NLO,old}+ \Delta \hat{H}_\text{eff,ss}^{\text{NLO}}
		\end{equation}
	one thus finds the relation	
	
		\begin{equation}
		\label{eq:all_together}
			\hat{H}_\text{eff,ss}^{\text{NLO,old}}  + \Delta \hat{H}_\text{eff,ss}^{\text{NLO}} \equiv  \hat{H}_{\text{ss}}^{\text{NLO}\prime \prime \prime}+ \frac{\nu}{c^2} \hat{H}_{\text{EOB,o}}^\text{N}\hat{H}_{\text{EOB,ss}}^{\text{LO}}.
		\end{equation}
	Remember that $\hat{H}_{\text{ss}}^{\text{NLO}\prime \prime \prime}$ is determined up to some free gauge parameters associated to the generating function $\hat{G}_{\text{ss}}^{\text{NLO}}$.
	Clearly, the choice of $\hat{G}_{\text{ss}}^{\text{NLO}}$ uniquely defines $\Delta \hat{H}_\text{eff,ss}^{\text{NLO}}$.
	For a better understanding, we place the terms that are not yet fixed on the left hand side of the equation:
	
		\begin{alignat}{1}
		\label{eq:all_together_2}
		  	\Delta \hat{H}_\text{eff,ss}^{\text{NLO}} - \big \{\hat{G}_{\text{ss}}^{\text{NLO}},\hat{H}_{\text{o}}^\text{N} \big \} =&  \hat{H}_{\text{ss}}^{\text{NLO}\prime \prime} -\hat{H}_\text{eff,ss}^{\text{NLO,old}} \nonumber\\
				+ & \frac{\nu}{c^2} \hat{H}_{\text{EOB,o}}^\text{N}\hat{H}_{\text{EOB,ss}}^{\text{LO}}.
		\end{alignat}	
	We recall that the EOB dynamics can be explicitly written as a deformation, in the ``small'' parameter $\nu$, of the well-known dynamics of a test particle in the Schwarzschild metric (for non spinning systems) or in the  Kerr metric (for spinning systems).
	In order to preserve this central feature, $ \Delta \hat{H}_\text{eff,ss}^{\text{NLO}}$ must thus vanish for $\nu = 0$.
	A straightforward calculation shows that this is satisfied if

		\begin{equation}
		\label{eq:Kerr_gen_func}
 			\hat{G}_{\text{ss}}^\text{NLO} = \frac{1}{c^6 r^2}\bigg [ -\frac{1}{2} \left(\bm{\chi_1}^2+(\bm{n}\cdot \bm{\chi_1})^2 \right)(\bm{n} \cdot \bm{p}) +  (\bm{p}\cdot \bm{\chi_1} )(\bm{n}\cdot \bm{\chi_1}) \bigg ],
		 \end{equation}
	where $\chi_1$ denotes the (dimensionless) spin of the largest body, i.e. of the Kerr black hole.
	$\hat{G}_{\text{ss}}^{3PN}$ is not uniquely defined, since it can contain arbitrary terms that vanish in the Kerr limit.
	The existence of this canonical transformation is not surprising.
	Indeed, we expect the dynamics of the Kerr metric, when expanded in PN orders, to be equivalent to the test-mass limit of the ADM Hamiltonian (\ref{eq:ADM_Ham}).
	Notice that the effects of the smaller spin are of order $\mathcal{O}(\nu^2)$, and are thus completely suppressed in the limit $\nu \to 0$.
	
	At this point, we turn the discussion to the special case of equatorial orbits and aligned spins.
	This is simply done by inserting the conditions $(\bm{\chi}_a \cdot \bm{p}) = (\bm{\chi}_a \cdot \bm{n})=0$ into Eq.~(\ref{eq:all_together_2}).
	The consistency of this simplification is discussed in the Appendix.
	The generating function $\hat{G}_{\text{ss}}^\text{NLO}$ takes the general form
		
		\begin{alignat}{2}
 			 \hat{G}_{\text{ss,al}}^{\text{NLO}}=  \frac{1}{c^6\, r}\bigg\{  & \bigg [ &&  \alpha_{11} \, \bm{p}^2(\bm{n}\cdot \bm{p})+ \beta_{11} (\bm{n}\cdot \bm{p})^3 \nonumber\\
					& &&+\left(\gamma_{11}-\frac{1}{2}\right) \frac{(\bm{n}\cdot \bm{p})}{r} \bigg]  \bm{\chi}_1^2\nonumber\\
			 	+ &  \bigg [ && \alpha_{12} \, \bm{p}^2(\bm{n}\cdot \bm{p}) + \beta_{12} (\bm{n}\cdot \bm{p})^3 \nonumber\\
					& && +\gamma_{12} \frac{(\bm{n}\cdot \bm{p})}{r} \bigg]  (\bm{\chi}_1\cdot \bm{\chi}_2)\nonumber\\
				+ &  \bigg [&& \alpha_{22} \, \bm{p}^2(\bm{n}\cdot \bm{p})  +  \beta_{22} (\bm{n}\cdot \bm{p})^3\nonumber\\ 
					& && +\left(\gamma_{22}-\frac{1}{2}\right) \frac{(\bm{n}\cdot \bm{p})}{r}\bigg]  \bm{\chi}_2^2\bigg\},
		\end{alignat}
	where the free gauge parameters $\alpha_{ab}(\nu)$, $\beta_{ab}(\nu)$ and $\gamma_{ab}(\nu)$ must vanish for $\nu = 0$.
	Notice that, in order to guarantee a symmetric treatment of both spins, we have introduced a term of $-1/2$ to the $\bm{\chi}_2^2$-dependent part of the generating function too.
	Eq.~(\ref{eq:all_together_2}) is solved by
\begingroup
\allowdisplaybreaks
	\begin{widetext}
		\begin{alignat}{2}
			\Delta \hat{H}_{\text {eff,ss}}^{\text{NLO}} =
				& \quad \frac{1}{c^6} \bigg [ &&  \alpha_{11}\frac{\bm{p}^4}{r^2} - 4\beta_{11} \frac{(\bm{n}\cdot \bm{p})^4}{r^2}+ (3\beta_{11}-2\alpha_{11}) \frac{(\bm{n}\cdot \bm{p})^2\, \bm{p}^2}{r^2} \nonumber\\
					& &&+ \frac{1}{4}\Bigg(-2 + 4\gamma_{11} - 4\alpha_{11} + 5\nu^2 +2\nu^3 + \frac{m_1}{m_2}( 7\nu^2 + 4\nu^3) + \left(\frac{m_1}{m_2}\right)^2(2\nu^2 +2\nu^3) \Bigg) \frac{\bm{p}^2}{r^3}\nonumber\\
					& &&+\frac{1}{8}\Bigg( 12 - 24\gamma_{11} -24\beta_{11} -16\alpha_{11} -12\nu^2 +12\nu^3 + \frac{m_1}{m_2}(3\nu -36\nu^2 +24\nu^3)\nonumber\\
						& && \quad+ \left(\frac{m_1}{m_2}\right)^2(-15\nu^2+12\nu^3)\Bigg) \frac{(\bm{n}\cdot \bm{p})^2}{r^3}\nonumber\\
					& && + \frac{1}{4}\left(2 - 4\gamma_{11} -12\nu^2 -\nu^3 +\frac{m_1}{m_2}(12\nu -26\nu^2-2\nu^3)+\left(\frac{m_1}{m_2}\right)^2(-14\nu^2-\nu^3)\right)\frac{1}{r^4} \bigg ] \bm{\chi}_1^2\nonumber\\
				& + \frac{1}{c^6} \bigg [ &&  \alpha_{22}\frac{\bm{p}^4}{r^2} - 4\beta_{22} \frac{(\bm{n}\cdot \bm{p})^4}{r^2}+ (3\beta_{22}-2\alpha_{22}) \frac{(\bm{n}\cdot \bm{p})^2\, \bm{p}^2}{r^2}\nonumber\\
					& &&+ \frac{1}{4}\Bigg(-2 + 4\gamma_{22} - 4\alpha_{22} + 5\nu^2 +2\nu^3 + \frac{m_2}{m_1}( 7\nu^2 + 4\nu^3) + \left(\frac{m_2}{m_1}\right)^2(2\nu^2 +2\nu^3) \Bigg) \frac{\bm{p}^2}{r^3}\nonumber\\
					& &&+\frac{1}{8}\Bigg( 12 - 24\gamma_{22} -24\beta_{22} -16\alpha_{22} -12\nu^2 +12\nu^3 + \frac{m_2}{m_1}(3\nu -36\nu^2 +24\nu^3)\nonumber\\
						& && \quad+ \left(\frac{m_2}{m_1}\right)^2(-15\nu^2+12\nu^3)\Bigg) \frac{(\bm{n}\cdot \bm{p})^2}{r^3}\nonumber\\
					& && + \frac{1}{4}\left(2 - 4\gamma_{22} -12\nu^2 -\nu^3 +\frac{m_2}{m_1}(12\nu -26\nu^2-2\nu^3)+\left(\frac{m_2}{m_1}\right)^2(-14\nu^2-\nu^3)\right)\frac{1}{r^4} \bigg ] \bm{\chi}_2^2\nonumber\\
				& + \frac{1}{c^6} \bigg [ &&  \alpha_{12}\frac{\bm{p}^4}{r^2} -4\beta_{12} \frac{(\bm{n}\cdot \bm{p})^4}{r^2} +(3\beta_{12}-2\alpha_{12})\frac{(\bm{n}\cdot \bm{p})^2\bm{p}^2}{r^2}\nonumber\\
					& &&+ \left(\gamma_{12} -\alpha_{12} + \nu^2 + 2\nu^3 + \left(\frac{m_1}{m_2}+\frac{m_2}{m_1}\right)(\nu^2+\nu^3)\right) \frac{\bm{p}^2}{r^3}\nonumber\\
					& &&+\frac{1}{4}\left( -12\gamma_{12} -12\beta_{12}-8\alpha_{12} -3\nu^2+24\nu^3+\left(\frac{m_1}{m_2}+\frac{m_2}{m_1}\right)12\nu^3\right) \frac{(\bm{n}\cdot \bm{p})^2}{r^3}\nonumber\\
					& && \left( -\gamma_{12}+6\nu -12\nu^2-\nu^3+ \left(\frac{m_1}{m_2}+\frac{m_2}{m_1}\right)\left(-6\nu^2-\frac{\nu^3}{2}\right)\right)\frac{1}{r^4} \bigg ] (\bm{\chi}_1 \cdot \bm{\chi}_2). 
		\end{alignat}
	\end{widetext}
\endgroup
	The simplest way of including these terms may be to add them to the whole effective Hamiltonian,

		\begin{equation}
		\label{eq:add_DeltaH}
			\hat{H}_{\text{eff}}^\text{old} \to \hat{H}_\text{eff} \equiv \hat{H}_{\text{eff,old}}^\text{old} + \Delta \hat{H}_\text{eff,ss}^\text{NLO}. 
		\end{equation}
	Of course, adding PN terms to EOB Hamiltonians can eventually lead to bad behaviors in the phase space region where the PN expansion fails, but the additional degrees of freedom given by the gauge parameters $\alpha_{ab}$, $\beta_{ab}$ and $\gamma_{ab}$ can in principle be used to calibrate the model.
	In this case, one would get something similar to Ref.~\cite{pan:10}, where an adjustable NLO spin-spin term $\propto \, (\nu \, S^2) / r^4$  was added to the effective Hamiltonian for calibration purposes. 
	However, we do not believe an inclusion of the new terms according to (\ref{eq:add_DeltaH}) to be satisfying.
	First of all, this would break the Kerr-like structure of the effective Hamiltonian.
	Secondly, the treatment of the spin would be made in a very different and non-straightforward way than in Refs.~\cite{dam:01,dam:08,nag:11}.
	Thirdly, one can verify that the existence of an ISCO would not be preserved for all choices of the gauge parameters.	
	For these reasons, we propose another approach.
	Instead of adding a new term to the effective Hamiltonian, we try to redefine the effective squared spin parameter $\chi_0^2$ entering the deformed Kerr metric, adding an appropriate NLO term.
	By contrast, we leave unmodified the ``linear'' spin $\chi_0$ appearing in the metric element $g_\text{eff}^{t\varphi}$ (\ref{eq:g^{tphi}}).
	Notice, in passing, that the introduction of the ``test spin'' $\sigma$ in Ref.~\cite{dam:08} is equivalent to a redefinition of the ``linear'' spin $\chi_0$ in $g_\text{eff}^{t\varphi}$, while leaving all squared spins untouched. 
	For this reason, the spin modification we are proposing is a very natural continuation of this philosophy.
	We replace all squared spins $\chi_0^2$ entering $g_\text{eff}$ (\ref{eq:metric}) according to
		
		\vskip-0.1in
		\begin{equation}
			\chi_0^2 \to \overline{\chi_\text{eff}^2} \equiv \chi_0^2 + \Delta \chi_\text{eff}^2,
		\end{equation}
	where
		\begin{alignat}{1}
			\Delta \chi_\text{eff}^2 \equiv & \frac{1}{c^2}\bigg[\left( a_{11}\bm{p}^2 +b_{11}(\bm{n}\cdot \bm{p})^2 + \frac{c_{11}}{r}\right)\chi_1^2\nonumber\\
				& \quad +\left( a_{22}\bm{p}^2 +b_{22}(\bm{n}\cdot \bm{p})^2 + \frac{c_{22}}{r}\right)\chi_2^2\nonumber\\
				& \quad +\left( a_{12}\bm{p}^2 +b_{12}(\bm{n}\cdot \bm{p})^2 + \frac{c_{12}}{r}\right)\chi_1 \chi_2\bigg]\nonumber\\
				& + \Delta \chi_\text{eff,NNLO}^2.			
		\end{alignat}

	The (yet undetermined) term $\Delta \chi_\text{eff,NNLO}^2$ has been inserted for calibration purposes in order to avoid bad behaviors of the new effective squared spin $\overline{\chi_\text{eff}^2}$ (see Sec.~\ref{sec:disc}).
	The function $\Delta_t$ becomes

		\begin{equation}
		\label{eq:Delta_t_new}
			\Delta_t = M^2\,r^2 P^n_m \Big[ A(u) + u^2 \overline{\chi_\text{eff}^2}(r, \bm{p}^2, \bm{n}\cdot \bm{p}) \Big],
		\end{equation}
	where the variable $u$ has to be set equal to $ c^{-2} r^{-1}$ only after Pad\'eing. 
	This ensures that the radial variable entering the effective squared spin does not gets resummed too (else, one would ``break'' the spin as a whole, treating the variables it depends on in different ways). 
	Of course, $\Delta_R$ has to be modified correspondingly (\ref{eq:D^-1}).

	The new effective squared spin, together with (\ref{eq:H_{eff,ss}^{LO}}), gives rise to the additional contribution

		\begin{equation}
				\Delta \hat{H}_\text{eff,ss}^\text{NLO} =\; \frac{1}{c^6\, r^2} \left( (\bm{n}\cdot \bm{p})^2 - \frac{\bm{p}^2}{2}\right) \Delta \chi_\text{eff}^2.
		\end{equation}
	A correct choice of the coefficients $a_{ab}$, $b_{ab}$, $c_{ab}$ and of the gauge parameters $\alpha_{ab}$, $\beta_{ab}$, $\gamma_{ab}$ solves Eq.~(\ref{eq:all_together_2}).
	The result is
	
	\begin{widetext}
		\begin{subequations}
			 \begin{alignat}{1}
				a_{11} = &\, \frac{\nu}{16}\left(-32\nu -22\nu^2 +\frac{m_1}{m_2}(21 -44\nu -44\nu^2)+ \left(\frac{m_1}{m_2}\right)^2(-21\nu -22\nu^2)\right)\\
				b_{11} = &\, 0\\
				c_{11} = &\, \frac{\nu}{16}\left( 88\nu +14\nu^2 +\frac{m_1}{m_2}(-117 + 196\nu +28\nu^2)+ \left(\frac{m_1}{m_2}\right)^2(117\nu +14\nu^2)\right)\\
				a_{12} = &\, \frac{\nu}{8}\left(24 - 53\nu -44\nu^2 + \left(\frac{m_1}{m_2}+\frac{m_2}{m_1}\right)(-32\nu-22\nu^2)\right)\\
				b_{12} = &\, 0\\
				c_{12} = &\, \frac{\nu}{8}\left( -120 +229\nu +28\nu^2 + \left(\frac{m_1}{m_2}+\frac{m_2}{m_1}\right)(112\nu+14\nu^2)\right).
			 \end{alignat}
		\end{subequations}
	\end{widetext}
	The coefficients $a_{22}$, $b_{22}$ and $c_{22}$ can be obtained from $a_{11}$, $b_{11}$ and $c_{11}$, just exchanging the particle label 1 and 2.
	The gauge coefficients are
	
	\begin{widetext}
		\begin{subequations}
			 \begin{alignat}{1}
				\alpha_{11} = &\, \frac{\nu}{32}\left( 32\nu +22\nu^2 + \frac{m_1}{m_2}(-21 + 44\nu +44\nu^2) + \left(\frac{m_1}{m_2}\right)^2(21\nu +22\nu^2)\right)\\ 
				\beta_{11} = &\, 0\\
				\gamma_{11} = &\, \frac{1}{4}\left( 2 -12\nu^2 -\nu^3 +\frac{m_1}{m_2}(12\nu -26\nu^2-2\nu^3)+ \left(\frac{m_1}{m_2}\right)^2(-14\nu^2 -\nu^3)\right)\\ 
				\alpha_{12} = &\, \frac{\nu}{16}\left(-24 +53\nu +44\nu^2 + \left(\frac{m_1}{m_2}+\frac{m_2}{m_1}\right)(32\nu+22\nu^2)\right)\\
				\beta_{12} = &\, 0\\
				\gamma_{12} = &\, \frac{\nu}{2}\left(12-24\nu -2\nu^2 + \left(\frac{m_1}{m_2}+\frac{m_2}{m_1}\right)(-12\nu -\nu^2)\right). 
			 \end{alignat}
		\end{subequations}
	\end{widetext}
	It is remarkable that all $b_{ab}$ and $\beta_{ab}$ vanish, thereby eliminating one third of the newly involved coefficients.	
	
		\begin{figure}[h!]
			\includegraphics[width=0.43\textwidth]{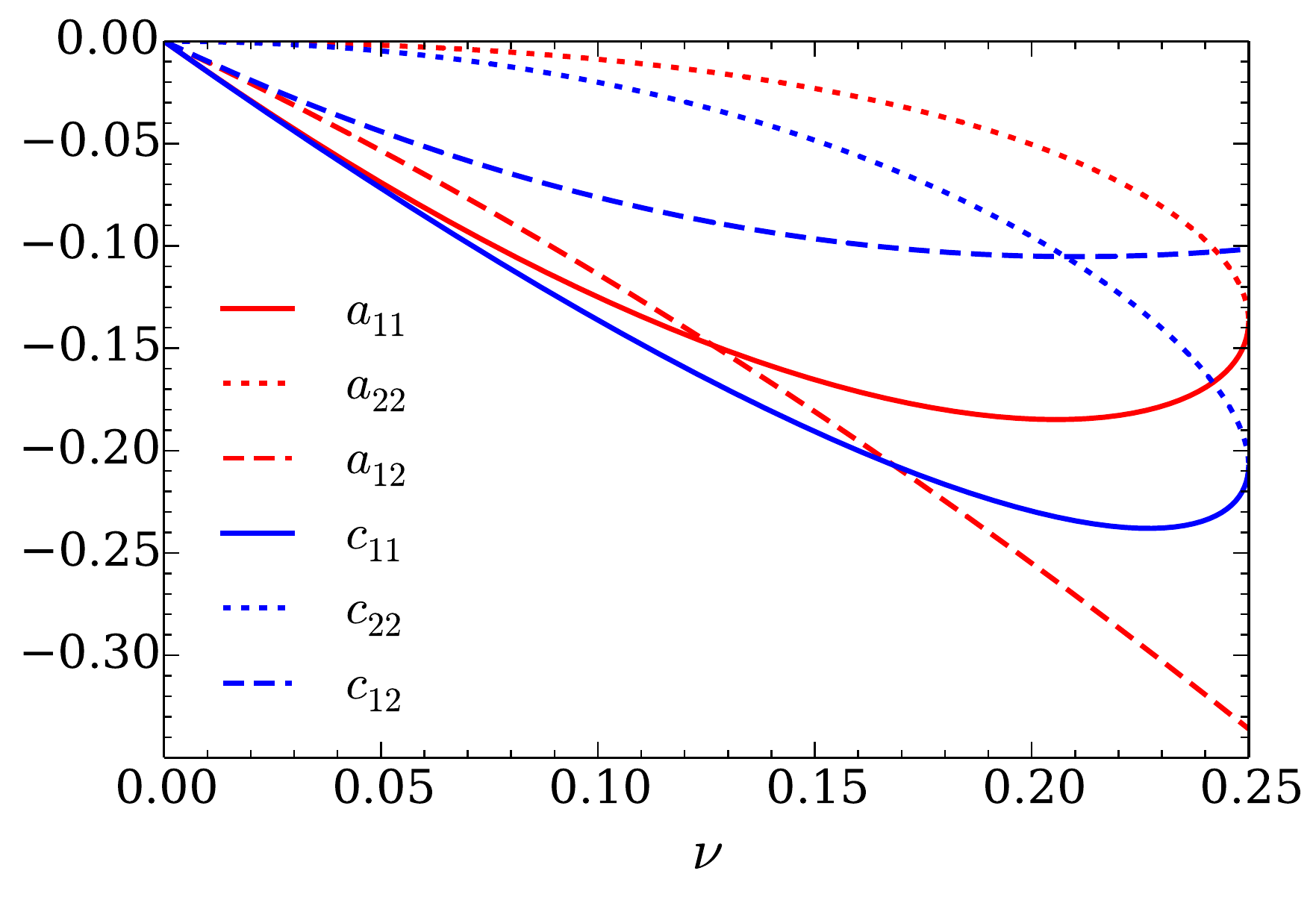}
			\caption{Plot of the coefficients $a_{ab}$ and $c_{ab}$ as a function of $\nu$.}
			\label{spin_coeff}
		\end{figure}
\section{Discussion}
\label{sec:disc}
	In this section, the consistency and some properties of the proposed EOB Hamiltonian are discussed.
	
	At first, it is easy to check that for $\nu \to 0$ all $a_{ab}$, $b_{ab}$ and $c_{ab}$ vanish. 
	Because of the notation involving the non-symmetric ratios $m_1/m_2$ and $m_2/m_1$, the $\nu$-dependence is not very explicit.
	However, the above statement follows from the simple fact that $\lim_{\nu \to 0}\nu \,m_1/m_2= 1$.
	The Kerr limit 
	
		\begin{equation}
			H_\text{EOB} \to H_\text{Kerr} \quad \text{for } \nu \to 0
		\end{equation}
	is therefore still valid, consistently with the usual interpretation of the EOB Hamiltonian as a $\nu$-dependent deformation of the Kerr (or Schwarzschild) metric.
	
	In this section, we limit the discussion to the equal masses ($m_1 = m_2$) and equal, aligned spins ($\bm{\chi}_1 = \bm{\chi}_2 = \bm{\chi}_0$) case. 
	Indeed,  we expect the most relevant discrepancies from the Kerr case to occur for both mass and spin ratios of the order of one, and thus we believe this particular choice of parameters to be representative for the whole ``non-Kerr'' behavior of the EOB dynamics. 

	Since $\bm{\chi}_0$ can be either aligned or anti-aligned with the angular momentum $\bm{l}$, we use the notation $\chi_0 \equiv |\bm{l}|^{-1}\,\bm{l} \cdot \bm{\chi}_0.$ 
	We take into account the spin-orbit effects up to NNLO, using the gyro-gravitomagnetic ratios as given by Eqs.~(55)-(56) of Ref.~\cite{nag:11}.
	Moreover, we calculate $\Delta_t$ with the Pad\'e approximant $P^1_3$ (\ref{eq:Delta_t_new}) and include the orbital dynamics up to 3PN (\ref{eq:AD^-1}).
	A numerical implementation of the new Hamiltonian shows that the existence of an innermost stable circular orbit (ISCO) is preserved.
	This is \textit{a priori} not obvious, since the strong-field properties are significantly influenced by adding non resummed PN terms.
		\begin{figure}
		\vspace{4pt}
 		   	\includegraphics[width=0.43\textwidth]{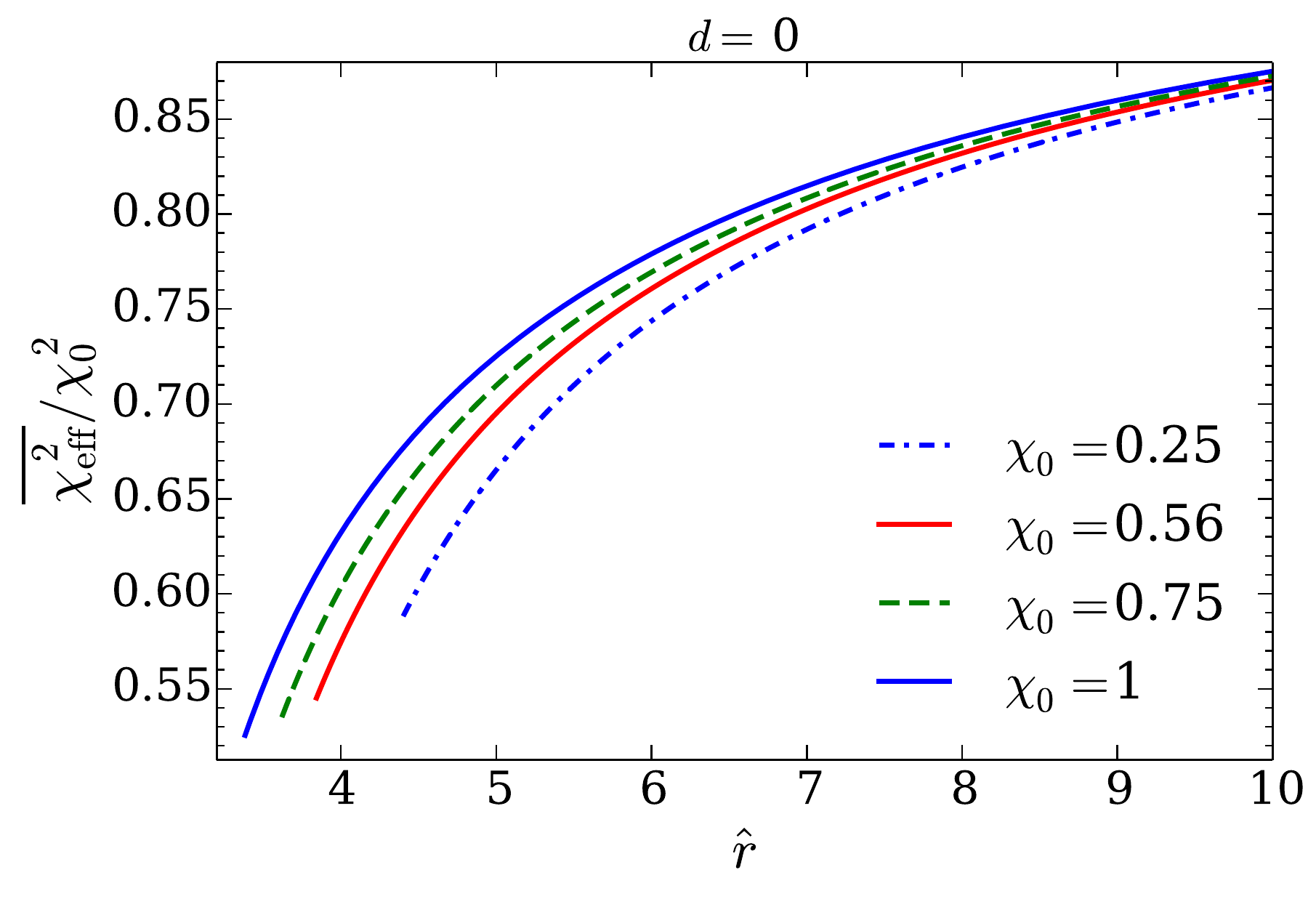}	
 		   	\includegraphics[width=0.43\textwidth]{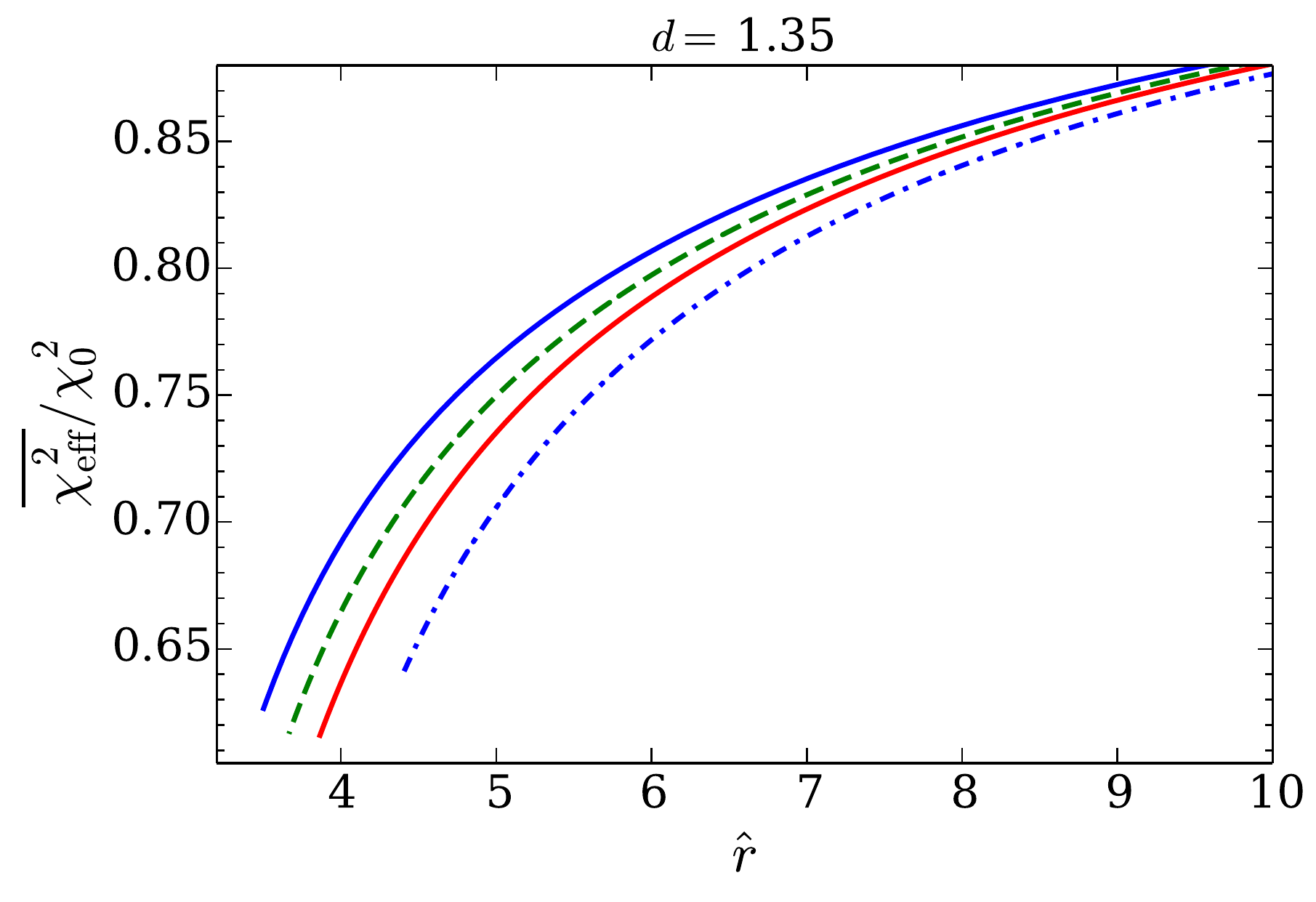}	
 		   	\includegraphics[width=0.43\textwidth]{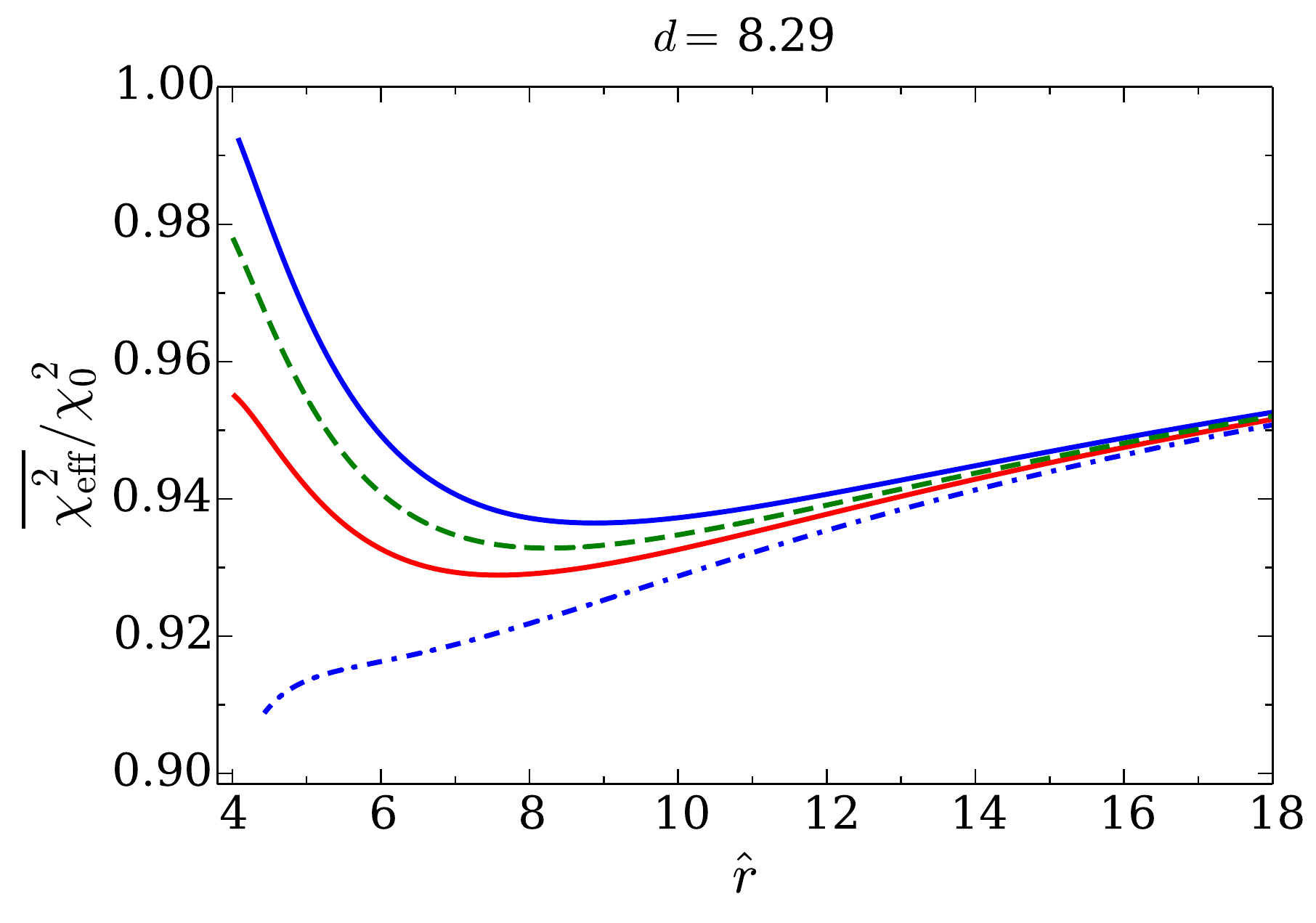}	
			\caption{Effective squared spin $\overline{\chi_\text{eff}^2}/ \chi_0^2$ for circular orbits plotted as a function of the dimensionless orbital radius $\hat{r} = c^2\,r$.
				Each line stops at its corresponding ISCO.}
			\label{fig:spin_circ}
		\end{figure}	  

	Before doing explicit calculations, however, it is necessary to fix the NNLO term $\Delta \chi_\text{eff,NNLO}^2$.
	The motivation of inserting it lies in the fact that, while the original effective spin $\chi_0$ has the great advantage of preserving the Kerr Bound $|\chi_0| \leq 1$ by construction, this is not necessarily true for the modified $\overline{\chi_\text{eff}^2}$ anymore, because of its dependence on the dynamical variables.
	Moreover, $\overline{\chi_\text{eff}^2}$ is not the square of any real function, and thus we also have to make sure that it never takes negative values. 
	In the far-field limit ($1/r \sim \bm{p}^2 \to 0 $), the original value $\overline{\chi_\text{eff}^2} \to \chi_0^2$ is recovered.
	By contrast, the strong-field behavior can eventually lead to inconsistencies such as the violation of the Kerr bound or a negative effective spin squared.
	The most natural thing to do is to correct these possible bad behaviors with an appropriate NNLO term. 
	Of course, an accurate determination of it can only be done by a comparison with numerical relativity or with the inclusion of higher order spin-spin terms.
	Here we just require consistency in the following, restricted way.
	With the simple ansatz

		\begin{equation}
			\Delta \chi_\text{eff,NNLO}^2  = \frac{d \, \nu}{c^4\,r^2}\big[\chi_1^2 + \chi_2^2+ \chi_1 \, \chi_2\big],
		\end{equation}
	we try to give a reasonable lower and upper bound for $d$ so that 
	
		\begin{equation}
	 		 0 \leq \overline{\chi_\text{eff}^2} \leq 1
		\end{equation}
	along stable and unstable circular orbits.
	A complete analysis of all orbits relevant for GW detection would of course require a numerical evaluation of a large portion of the  phase space.
	Anyway, we believe that restricting the analysis to circular orbits does not influence significantly the estimation of $d$.	
	Indeed, unstable circular orbits lie on the light ring, which corresponds to the smallest separation radius that can be reached by an eccentric orbit with given angular momentum. 
	Since $\bm{p}^2$ is expected to increase with $1/r$, and since $a_{ab}$ and $c_{ab}$ have the same sign (see Fig.~\ref{spin_coeff}), it is reasonable to think that the effective squared spin $\overline{\chi_\text{eff}^2}$ is most likely to assume an unphysical value when the separation radius lies on the light ring. 

	The dynamics of circular orbits is obtained setting the radial momentum $(\bm{n}\cdot \bm{p}) = 0$ and solving

		\begin{equation}
			\label{eq:circ_orb}
			\frac{\partial}{\partial r}H_\text{EOB}(r,l) =0,
		\end{equation}
	where $l = r \sqrt{\bm{p}^2-(\bm{n}\cdot \bm{p})^2 }$ denotes the (rescaled) angular momentum.
	Among the solutions, the stable circular orbits correspond to the local minima of the effective potential $H_\text{EOB}(r,l)$, while the light ring $r_{LR}(l)$ corresponds to the local maxima. 
	The ISCO is located at the turning point, and is thus determined by the additional condition

		\begin{equation}
			\frac{\partial^2}{\partial r^2}H_\text{EOB}(r,l) = 0.
		\end{equation}	
	 
		\begin{figure}
			\includegraphics[width=0.43\textwidth]{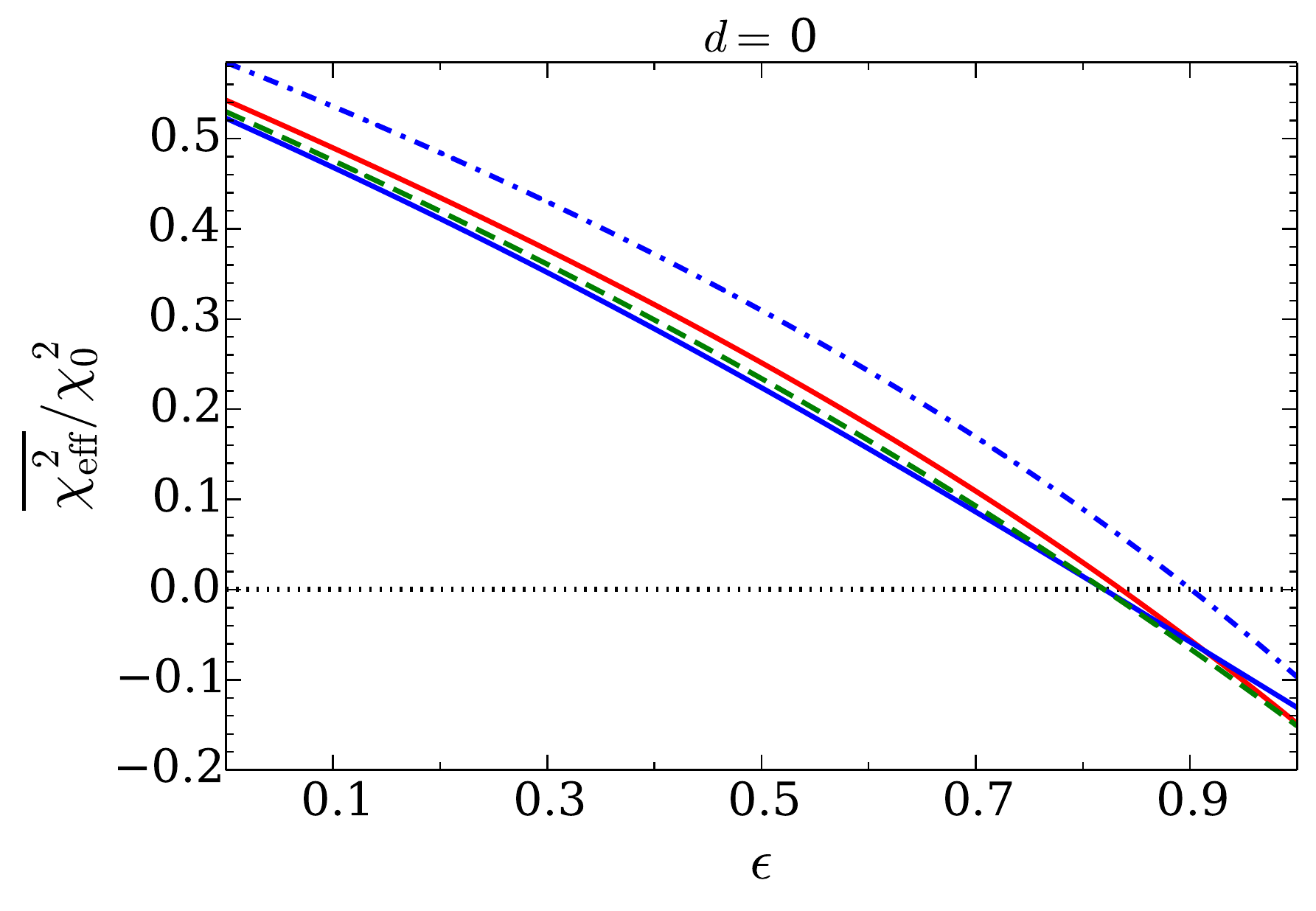}
			\includegraphics[width=0.43\textwidth]{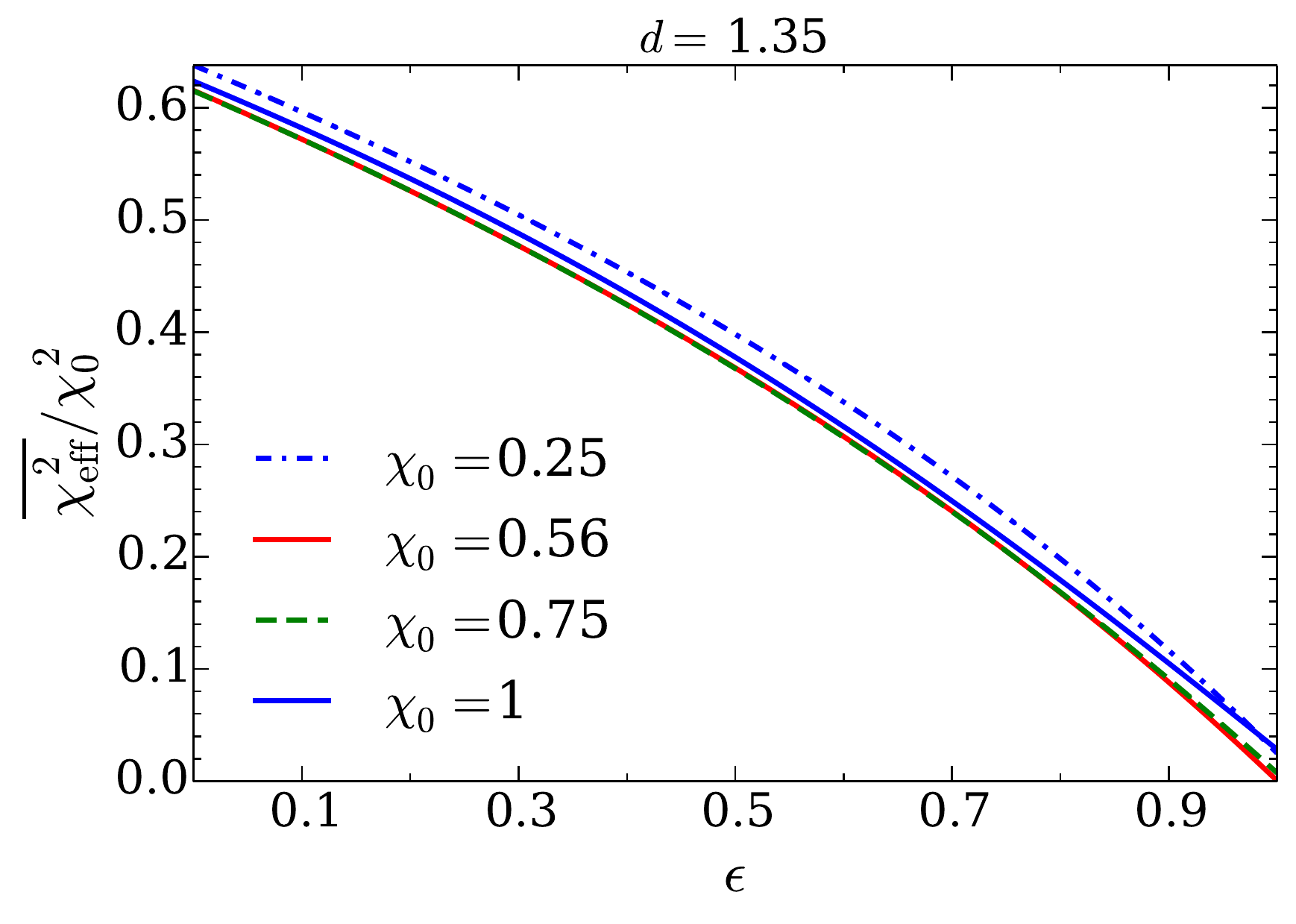}
			\includegraphics[width=0.43\textwidth]{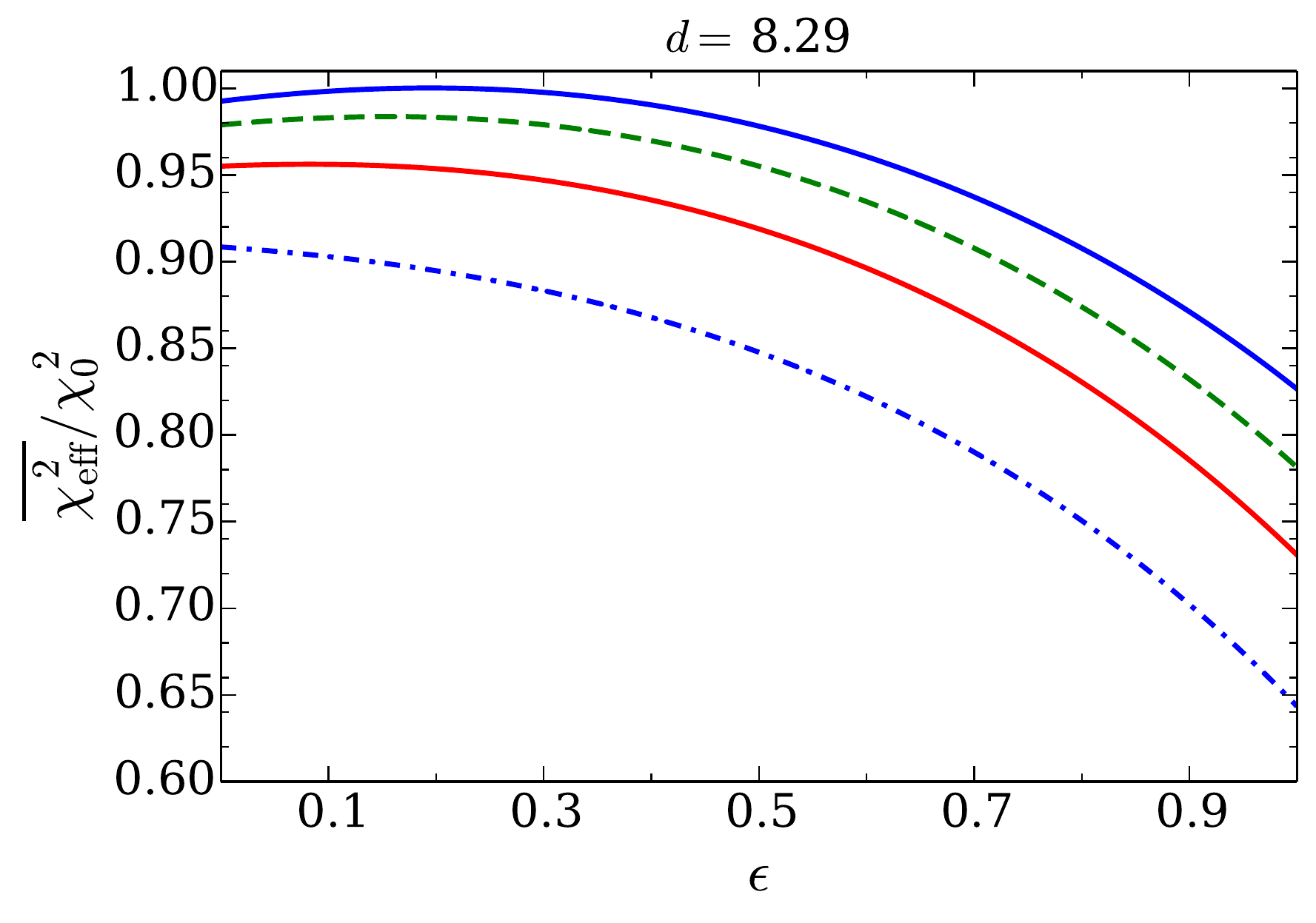}
			\caption{Effective squared spin $\overline{\chi_\text{eff}^2}/ \chi_0^2$ evaluated at the lightring $r_{LR}$ for eccentric orbits as a function of the eccentricity $\epsilon$.}
			\label{fig:spin_ecc}
		\end{figure}		

		\begin{figure}[h!]

			\includegraphics[width=0.43\textwidth]{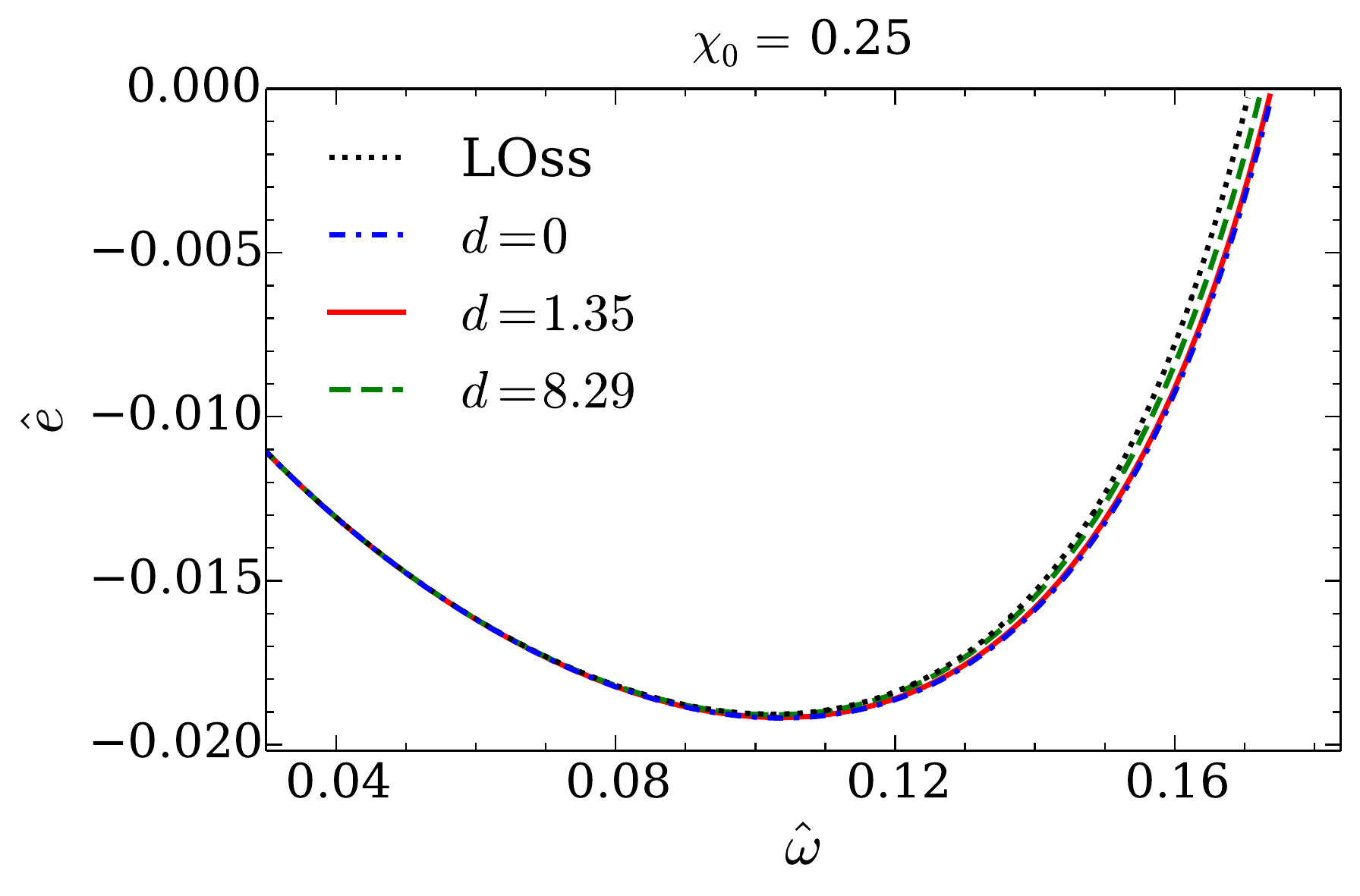}
                        \includegraphics[width=0.43\textwidth]{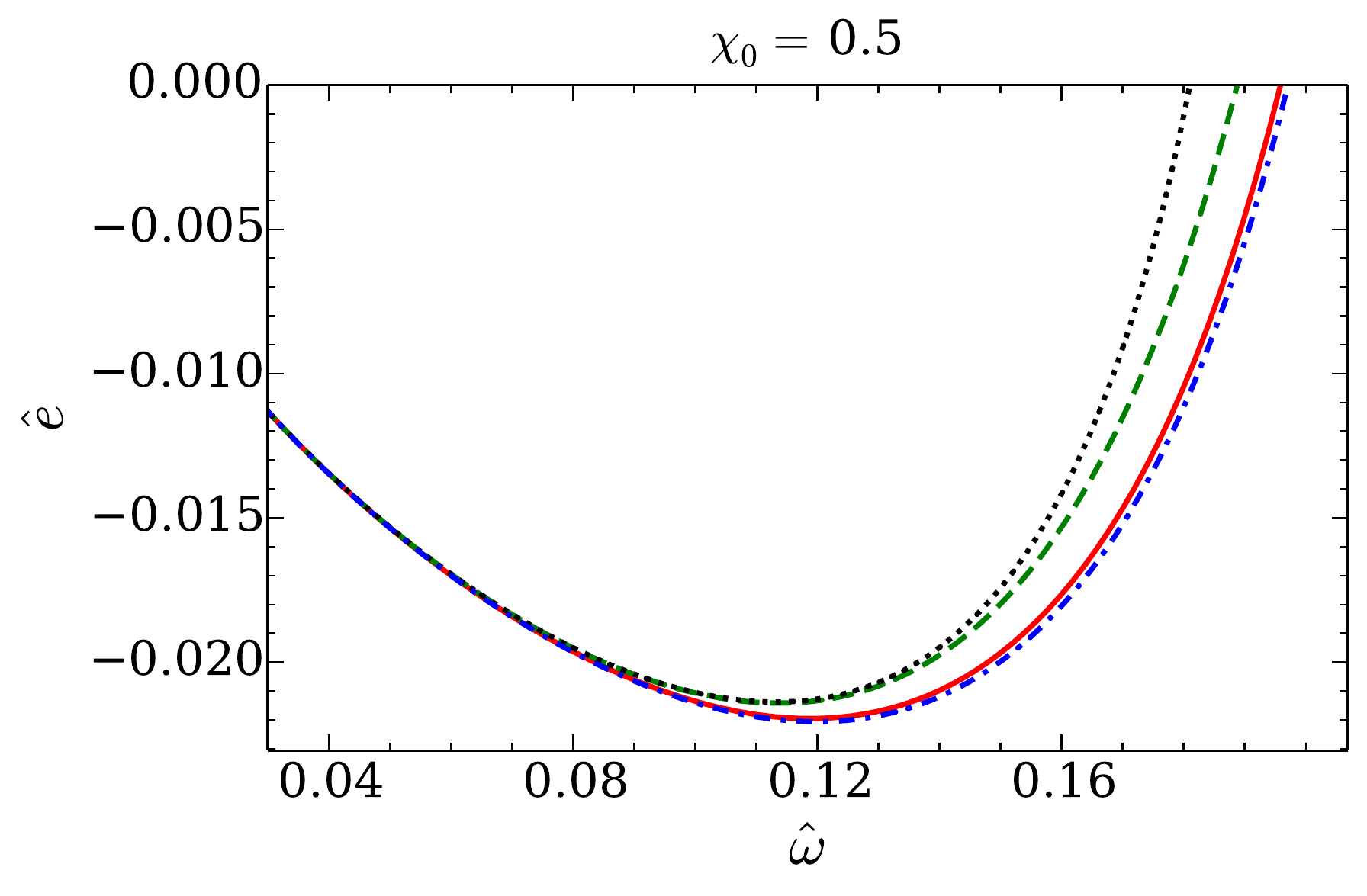}
	                \includegraphics[width=0.43\textwidth]{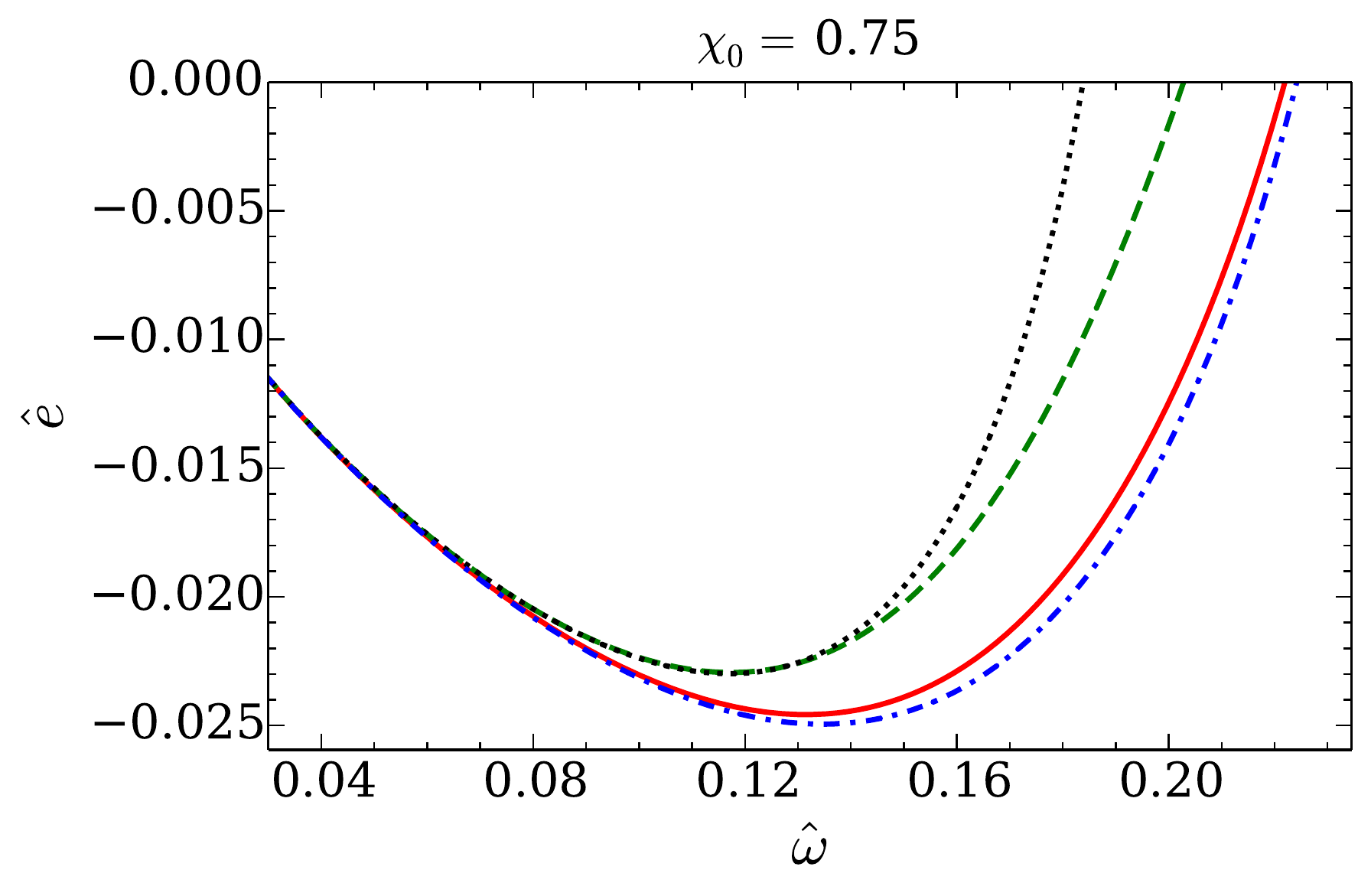}
                        \includegraphics[width=0.43\textwidth]{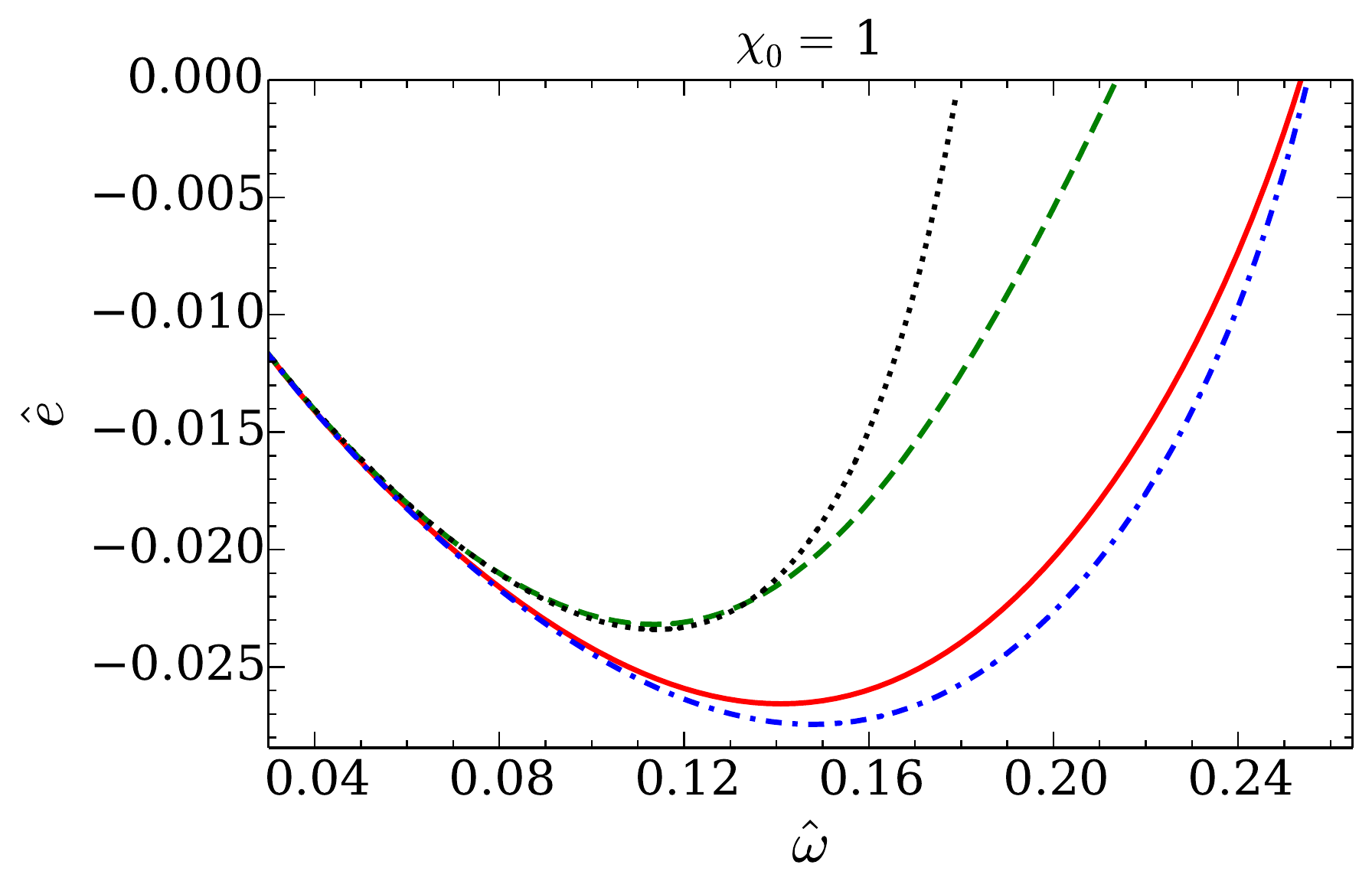}

			\caption{Binding energy curve for circular orbits for different values of $\chi_0$. 
			 	The black dotted line denotes the EOB Hamiltonian with LO spin-spin coupling and NNLO spin-orbit coupling, as given by Ref.~\cite{nag:11}.
			 	Equal masses and equal spins are assumed.}
			\label{fig:ecirc}
		\end{figure}
Evaluating the effective radial potential $ H_\text{EOB}(r,l)$ for $\chi_0$ ranging from $-1$ to $1$, we found the estimation $1.35 \lesssim d \lesssim 8.29$.  
	This means that, if we want $\overline{\chi_\text{eff}^2}$ to always be a positive number, the introduction of a non-zero $\Delta \chi_\text{eff,NNLO}^2$ is necessary.
	The effective squared spin $\overline{\chi_\text{eff}^2}$ is plotted in Fig.~\ref{fig:spin_circ} for stable circular orbits and in Fig.~\ref{fig:spin_ecc} in the correspondence of the light ring.
	There, the light ring is interpreted as the smallest separation radius of an orbit with eccentricity 

		\begin{equation}
			\epsilon = \frac{r_{max} - r_{LR}}{r_{max} + r_{LR}}, 
		\end{equation}
	where $r_{max}$ is the largest separation radius. 
	In addition to the limiting cases $d = 1.35$ and $d = 8.29$, we also consider the ``purely analytical'' case where the positivity requirement for $\overline{\chi_\text{eff}^2}$ is dropped off by removing the corrective term $\Delta \chi_\text{eff,NNLO}^2$, which is equivalent to set $d=0$.	
	As shown in Fig.~\ref{fig:spin_ecc}, the limiting value $\overline{\chi_\text{eff}^2} = 0$ is reached for $d \approx 1.35$, $\chi_0 \approx 0.56$ and $\epsilon \to  1$, while $\overline{\chi_\text{eff}^2} = 1$ for $d \approx 8.29$,  $\chi_0 = 1$ and $\epsilon \approx 0.2$.
	It is also visible that  $d =8.29$ is responsible for a non-monotonic dependency of $\overline{\chi_\text{eff}^2}$ on $r$ for large spin values.	
	On the other hand, setting $d=0$ does not actually seem an unreasonable choice, unless one is interested in highly eccentric ``zoom-whirl'' orbits: indeed, for all stable circular orbits  $\overline{\chi_\text{eff}^2}/ \chi_0^2 > 1/2$, while for eccentric orbits $\overline{\chi_\text{eff}^2} > 0$ up to $\epsilon \sim 0.8$.

	As in  Ref.~\cite{dam:08}, we also investigate the binding energy of the system along circular orbits.
	In Fig.~\ref{fig:ecirc}, the dimensionless, non relativistic energy
	
		\begin{equation}
			\hat{e} = \frac{H_\text{EOB}}{M\, c^2} -1 
		\end{equation}		
	is plotted as a function of the dimensionless frequency
	
		\begin{equation}
			\hat{\omega} = \frac{1}{c^3\,\mu}\frac{\partial}{\partial l}H_\text{EOB}. 
		\end{equation}		
	Notice that the ISCO corresponds to the curve minima.
	For comparison, we also show the binding energy for LO spin-spin coupling, according to Ref.~\cite{nag:11}.

	As already mentioned, the presence of larger spins aligned with the angular momentum is responsible for more bounded orbits.
	As shown in Fig.~\ref{fig:ecirc}, the inclusion of NLO spin-spin terms strengthens this effect, leading to a binding energy increase up to $\sim 15 \%$ (for the $\chi_0 = 1$ and $d =0$ case). 
	Notice that there is a subtlety which may be a source of confusion: since the effective Kerr parameter squared gets smaller after the NLO spin-spin inclusion ($\overline{\chi_\text{eff}^2} \leq \chi_0^2$, see Figs.~\ref{fig:spin_circ} and \ref{fig:spin_ecc}), one might have wrongly expected smaller binding energies. 
	Instead, this simply suggests that in the Kerr-like metric the relation ``a larger effective spin implies a more bounded orbit'' is essentially a spin-orbit feature, while the quadratic appearances of the Kerr parameter seem to act in the opposite way.
	
		\begin{figure}[h!]
			\includegraphics[width=0.43\textwidth]{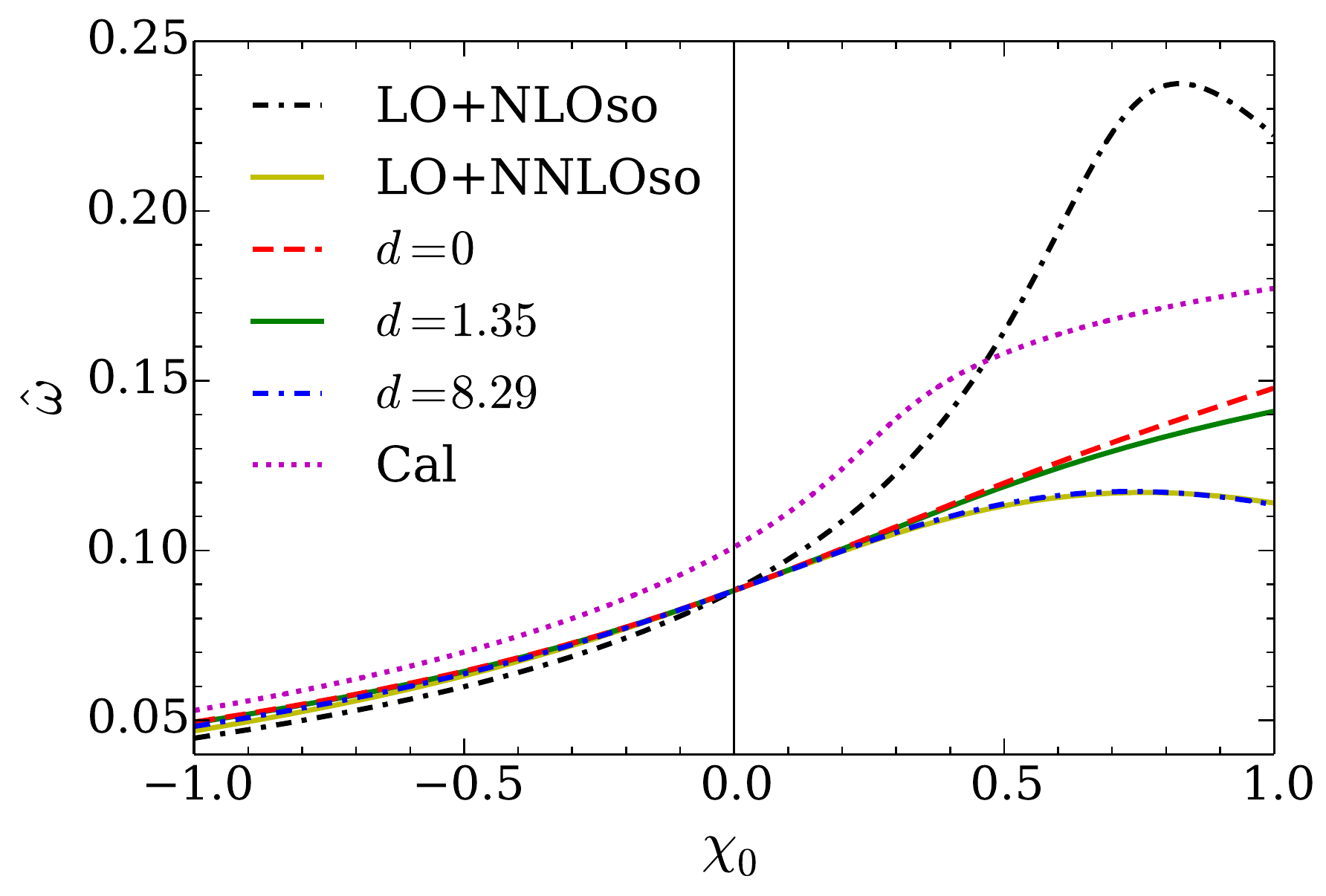}
			\caption{Frequency $\hat{\omega}$ at the ISCO as a function of $\chi_0$. Equal masses and equal spins are assumed.}
			\label{fig:freq_isco}
		\end{figure}	  
  	
	A second thing that can be observed from Fig.~\ref{fig:ecirc} is that the curve for the ```maximal'' $d \approx 8.29$ is very similar (at least in the range of stable orbits) to the curve with LO spin-spin effects.
	Thus, the actual role of the maximal $d$ becomes clear: restoring, in the ISCO region, an effective spin squared close to the LO one, $\chi_0^2$ (see also Fig.~\ref{fig:spin_circ}), it acts almost compensating the NLO spin-spin effects. 
	For this reason, we believe that a large $d$ may be a bad choice.
	On the other hand, a ``purely analytical'' implementation (i.e., with $d=0$) does not show any particular problem up to large eccentricities, and may therefore be a reasonable choice for an uncalibrated EOB model. 

	This argument is supported by Fig.~\ref{fig:freq_isco}, that shows (as in Ref.~\cite{bar:10}) the orbital frequency $\hat{\omega}_\text{ISCO}$ at the ISCO.
	The curve called ``LO+NLOso" corresponds to the prediction of Ref~\cite{dam:08}, and thus includes LO spin-spin effects and just NLO spin-orbit effects, while ``LO+NNLOso'' includes LO spin-spin and NNLO spin-orbits effects (according to Ref.~\cite{nag:11}).
	Spin-spin effects at NLO are implemented in the three lowest curves (that correspond to the model discussed up to now, and thus reproduce spin-orbit effects at NNLO accuracy).
	As a further comparison, we also have included the curve predicted by Ref.~\cite{tar:12}, that we denote by ``Cal".
	The corresponding EOB Hamiltonian is based on the model developed in Refs.~\cite{bar:09,bar:10,bar:11}.
	It includes spin-orbit effects at NNLO and spin-spin effects at LO, and has been calibrated inserting appropriate 3PN spin-spin and 4.5PN spin-orbit terms.
	``Cal'' generates reliable waveforms up to $\chi_0 \sim 0.7$.
	Because of a different resummation of the orbital part, the non spinning behavior differs quite significantly from our model.
	In particular, in ``Cal'' the functions $\Delta_t$ and $\Delta_R$ depend on an adjustable parameter $K$ \cite{bar:10}, which determines the radii where the horizons of the effective metric are located.
	The calibration of $K$  is responsible for the discrepancy at $\chi_0 = 0$  between  ``Cal'' and our model.
	This can be understood by looking at Fig.~1 of Ref.~\cite{bar:10}, that corresponds to  a choice of $K$ different from the one used in ``Cal''.
	For $\chi_0 = 0$, one reads a value $\omega_\text{ISCO} \approx 0.90$, which is quite close to the value $\omega_\text{ISCO} \approx 0.88$ predicted by our model.
	In the EOB Hamiltonian of this paper, there is no parameter analogous to $K$ that can be calibrated against numerical relativity.
	However, since $K$ appears in the expansion starting from 4PN, the discrepancy at $\chi_0 = 0$ might be reduced or even disappear when including higher order (and eventually calibrated) terms in the functions $\Delta_t$ and $\Delta_R$.
	We remark that the exact radial potential $A(u)$  is known at linear order in the symmetric mass-ratio \cite{bar:12}.
	Moreover, radial potentials at full 4PN \cite{bini:13}, or with some ``fiducial values'' up to 5PN (obtained through gravitational self-force calculations), see e.g. Ref.~\cite{ dam:13}, have already been inserted into an EOB model, however only for the non spinning case.

	Fig.~\ref{fig:freq_isco} shows that the system gets ``speeded up'' by the action of the NLO spin-spin coupling. 
	The most interesting point is that the the system is moved into the right direction (assumed to be the one shown by ``Cal''), especially for the purely analytical $d=0$ case. 	
	
\section{Conclusion}		
\label{sec:conclusion}
	We have shown that, by adding a term of fractional 1PN order to the effective Kerr parameter squared of the effective-one-body model developed in Refs.~\cite{dam:01,dam:08,nag:11}, it is possible to reproduce the  next-to-leading order, spin-spin contribution of the PN expanded Hamiltonian for two black holes with spins aligned with the angular momentum. 	
	In particular, this is possible thanks to a specific canonical transformation quadratic in the spins that has to be added to all transformations already found in the above references.	
	The additional spin-squared term vanishes whenever the mass-ratio tends to zero, so as to correctly reproduce the exact Kerr dynamics.	

	We have then evaluated the dynamics of circular orbits in the case of equal masses and spins.
	As a significant result, the effective radial potential still preserves the usual structure, reproducing local minima and maxima (corresponding to stable and unstable orbits, respectively) and also showing the existence of an ISCO.
	We recall that the location of the ISCO is of particular relevance for GW detection, since it describes the amount of energy that has been released during the inspiralling. 

	The general effect of the additional terms is to reduce the effective Kerr parameter squared.
	The problem is that, in the strong-field region, it can even vanish or become negative, thus breaking e.g. the horizon structure of the effective metric.
	In order to avoid this, it is possible to further modify the effective Kerr parameter squared inserting an additional term of fractional 2PN order, that should possibly be calibrated with numerical relativity.
	We have proposed a simple radial-dependent ansatz for the term in question, giving an estimation for his coefficient in order to preserve not only the positivity, but also the Kerr bound.
	Such bad behaviors, however, only happen in the regime of highly eccentric orbits ($\epsilon \sim 0.8$). 
	In addition, a comparison with an EOB model calibrated with numerical relativity shows that the inclusion of NLO spin-spin terms leads to an improvement in the description of the frequency at the ISCO, which is most relevant right in the case where no bound-preserving NNLO additional terms have been inserted.
	For these reasons, we believe that an EOB model with just NLO spin-spin terms is sufficiently self-consistent. 

	In general, the effect of the NLO spin-spin coupling is that of increasing the binding energy and the frequency at the ISCO.
	The need for further improvements still remains, yet it is clear that the inclusion of NLO spin-spin effects points in the right direction. 

\appendix
\setcounter{secnumdepth}{0}
\renewcommand{\theequation}{A.\arabic{equation}}

\section{Appendix: Existence of equatorial orbits}
\label{sec:existence}
	In this section we want to briefly motivate that constraining the orbital evolution to the equatorial plane, while holding the spins $\bm{S}_1$ and $\bm{S}_2$ fixed along the $e_3$ direction, is consistent with the conservative dynamics up to NLO in the spin-spin coupling.
	It has already been proved \cite{tessm:10} that the conservative 2.5PN dynamics of maximally rotating compact binaries does not allow the spins to precess, if they are both initially aligned with the total angular momentum $\bm{J}$.
	In Ref. \cite{tessm:10}, this statement is shown using an approach which dates back to Dirac.
	The idea is to express the parallelism of $\bm{S}_1$, $\bm{S}_2$ and $\bm{J}$ as a set of constraints 

		\begin{equation}
		\label{ }
			C_a(x,\,p,\,S)=0
		\end{equation}
	and  to show that their time derivative can be written in the form

		\begin{equation}
			\label{constr_cond}
			\dot{C}_a(x,\,p,\,S)= \sum_{b}D_{ab}(x,\,p,\,S)\,C_b.
		\end{equation}
	This implies that all time derivatives of the constraints are a linear combination of the constraints themselves, and thus vanish at an initial time $t=0$ when all $C_a$'s are set to zero.
	This in turn guarantees that the parallelism is conserved even at later times.
	Denoting the (rescaled) orbital angular momentum as

		\begin{equation}
		\label{ }
			\bm{l}= r\, \bm{n} \times \bm{p},
		\end{equation}
	the constraints can be written as

		\begin{equation}
		\label{ }
			C_a:=\bm{S}_a -\lambda_a \bm{l}= 0,
		\end{equation}
		
	where $\lambda_a :=|\bm{S}_a |\,|\bm{l}|^{-1}$. 
	Ref. \cite{tessm:10} shows that Eq.~(\ref{constr_cond}) is valid if one can express the time derivative $\dot{\bm{S}}_b$ of the spins as a linear combination of the constraints.
	The generalization to the NLO spin-spin coupling turns out to be straightforward. Without loss of generality, consider only the spin $\bm{S}_1$.
	One has to keep into account the two additional terms $\big\{ \bm{S}_1, H_{\text{S}_1^2}^{\text{NLO}}\big\}$ and $\big\{ \bm{S}_1, H_{\text{S}_1 \text{S}_2}^{\text{NLO}}\big\}$ appearing in its first time derivative.
	Formally, $H_{\text{S}_1^2}^{\text{NLO}}$ only contains terms of type $ A\, \bm{S}_1^2$ and $A_{vw}\, \left(\bm{S}_1\cdot \bm{v}\right)\left(\bm{S}_1\cdot \bm{w}\right)$, where the vectors $\bm{v}$ and $\bm{w}$ can either denote  $\bm{n}$ or $\bm{p}$.
	Analogously, the terms appearing in $H_{\text{S}_1 \text{S}_2}^{\text{NLO}}$ are of type $B\, \left(\bm{S}_1\cdot \bm{S}_2\right)$ and $B_{vw}\, \left(\bm{S}_1\cdot \bm{v}\right)\left(\bm{S}_2\cdot \bm{w}\right)$.
	Notice that the coefficients $A$, $A_{vw}$, $B$, and $B_{vw}$ are functions of $\bm{r}$ and $\bm{p}$, and are independent of the spins. The evaluation of the Poisson brackets leads to the following terms:

	\begin{widetext}	
		\begin{subequations}
			\begin{alignat}{2}
				\left\{ \bm{S}_1, A\, \bm{S}_1^2\right\}& = \,\,&& 0\\
				\left\{ \bm{S}_1, A_{vw}\, \left(\bm{S}_1\cdot \bm{v}\right)\left(\bm{S}_1\cdot \bm{w}\right)\right\} &=  && A_{vw} \big(\left(\bm{S}_1\cdot \bm{w}\right) \left( \bm{v} \times \bm{S}_1\right) + \left(\bm{S}_1\cdot \bm{v}\right) \left( \bm{w} \times \bm{S}_1\right) \big)\nonumber\\
					&=  && A_{vw} \big( [ \left(\bm{S}_1\cdot \bm{w}\right) - \lambda_1\left(\bm{l}\cdot \bm{w}\right)] \left( \bm{v} \times \bm{S}_1\right) \nonumber\\
			\label{eq:app_PB2}
						& && +  [\left(\bm{S}_1\cdot \bm{v}\right) -\lambda_1\left(\bm{l}\cdot \bm{v}\right)] \left( \bm{w} \times \bm{S}_1\right)\big)\\
				\left\{ \bm{S}_1, B\, \left(\bm{S}_1\cdot \bm{S}_2\right)\right\}&= &&B \left( \bm{S}_2 \times \bm{S}_1\right)\nonumber\\
			\label{eq:app_PB3}
					 & = && B [\left( \bm{S}_2 \times \bm{S}_1\right)-\lambda_1\left( \bm{S}_2 \times \bm{l}\right)]+B\lambda_1[\left( \bm{S}_2 \times \bm{l}\right)-\lambda_2 \left( \bm{l} \times \bm{l}\right)]\\
				\left \{ \bm{S}_1, B_{vw}\, \left(\bm{S}_1\cdot \bm{v}\right)\left(\bm{S}_2\cdot \bm{w}\right)\right\} &= && B_{vw}  \left(\bm{S}_2\cdot \bm{w}\right) \left( \bm{v} \times \bm{S}_1\right)\nonumber\\
			\label{eq:app_PB4}
					& = &&B_{vw}  [ \left(\bm{S}_2\cdot \bm{w}\right)-\lambda_2\left(\bm{l}\cdot \bm{w}\right)] \left( \bm{v} \times \bm{S}_1\right).
			\end{alignat}
		\end{subequations}
	\end{widetext}
	The last step of Eqs.~\eqref{eq:app_PB2}, \eqref{eq:app_PB3}, and \eqref{eq:app_PB4} uses the fact that, by construction, $\left(\bm{l}\cdot \bm{n}\right)=\left(\bm{l}\cdot \bm{p}\right)=0$.
	This, together with the result of Ref.~\cite{tessm:10}, shows that the  $\dot{\bm{S}}_b$ can be expressed as a linear combination of the constraints $C_a$, and therefore Eq.~(\ref{constr_cond}) also holds at NLO.

 	At last, we wish to argue that the alignment constraint is invariant under the transformation from ADM to EOB coordinates. 
	First of all, it is clear that all canonical transformations that do not involve spin variables preserve the alignment.
	Indeed, the only vectors building up their respective generating functions are the vectors $\bm{r}$ and $\bm{p}$, which are therefore transformed (according to Eq.~(\ref{eq:coord_transf})) into linear combinations of themselves, thereby remaining on the plane perpendicular to the spin.
	The same argument is valid for the canonical transformation $\hat{G}_\text{ss,al}^\text{NLO}$, since the spins variables appearing there just have scalar character.
	By contrast, this might not seem obvious in the case of $\hat{G}_\text{ss}^\text{LO}$, and thus we resolved to perform an explicit calculation.
	The transformation reads, after inserting the alignment constraint for $\bm{r}$ and $\bm{p}$:

		\begin{alignat}{2}
			\bm{r}' = & \,\, \bm{r}\bigg( 1 - \frac{1}{c^4} \frac{\bm{\chi}_0^2}{2r^2}\bigg)\\
			\bm{p}' = & \,\, \bm{p} + \frac{1}{2c^2\,r^2} \bigg[\bm{p}'\, \bm{\chi}_0^2 - 2\bm{n}\,(\bm{n} \cdot \bm{p}')\bm{\chi}_0^2\nonumber\\
				&+ \bm{\chi}_0 \,\Big((\bm{n} \times \bm{p}')\cdot (\bm{\chi}_0 \times \bm{n})\Big)\bigg].
		\end{alignat}
	The first equation already shows that $(\bm{r}'\cdot \bm{\chi}_0)=0$.
	Using the fact that $(\bm{n} \times \bm{p})\cdot (\bm{\chi}_0 \times \bm{n})=0$, we insert the second equation into the expression $(\bm{\chi}_0\times \bm{n})\cdot (\bm{n} \times \bm{p}')$ obtaining
	
		\begin{equation}
			(\bm{\chi}_0\times \bm{n})\cdot (\bm{n} \times \bm{p}')\left( 1 -\frac{1}{2c^4\,r^2}\left(\bm{\chi}_0^2- (\bm{\chi}_0\times \bm{n})^2\right) \right)=0,
		\end{equation}
	from which immediately follows that $(\bm{n} \times \bm{p}')\cdot (\bm{\chi}_0 \times \bm{n})=0$.
	This means that $\bm{p}'$ is again a linear combination of $\bm{r}$ and $\bm{p}$ and lies therefore on the plane perpendicular to $\bm{\chi}_0$.
	Very similar arguments can be used for the spin-orbit effects, as well as for their corresponding canonical transformation into EOB coordinates. 

\begin{acknowledgments}
We thank Thibault Damour for a very fruitful discussion, which helped to clarify many important points.
Moreover, we benefitted from discussions with Jan Steinhoff, Enrico Barausse and Alessandro Nagar.
S.B. is supported by the Swiss National Science Foundation. 
We thank the referee for useful comments.
\end{acknowledgments}

 \bibliography{./biblio}
 \bibliographystyle{apsrev}
 
\end{document}